\documentclass[longauth]{aa} 

\usepackage{graphicx}
\usepackage{txfonts}
\usepackage{amsmath}

\newcommand{\nh}{N_{\rm H}}

\begin{document}

   \title{A deep X-ray look to the most obscured quasar at z $\sim$ 3.6 and its environment}

   \subtitle{}

   \author{I. Villani
          \inst{1,2},
          L. Zappacosta
          \inst{2},
          E. Piconcelli
          \inst{2},
          M. Ginolfi
          \inst{3,4},
          F. Ricci
          \inst{1,2},
          F. La Franca 
          \inst{1,2},
          F. Arrigoni-Battaia
          \inst{5},
          A. Bongiorno
          \inst{2},
          S. Cantalupo
          \inst{6},
          S. Carniani
          \inst{7},
          F. Civano
          \inst{8},
          A. Comastri
          \inst{9},
          F. Fiore
          \inst{10,11},
          R. Maiolino
          \inst{12,13,14},
          L. Pentericci
          \inst{2},
          C. Ricci
          \inst{15,16},
          R. Schneider
          \inst{2,17,18,19},
          R. Valiante
          \inst{2,17},
          C. Vignali
          \inst{9,20},
          F. Vito
          \inst{9}
          }

\institute{Dipartimento di Matematica e Fisica, Università degli Studi Roma Tre, via della Vasca Navale 84, 00146 Roma, Italy
              \\
              \email{ilaria.villani@uniroma3.it}
         \and
             INAF – Osservatorio Astronomico di Roma, via di Frascati 33, 00078 Monte Porzio Catone, Italy
             \and
             INAF - Osservatorio Astrofisico di Arcetri, Largo E. Fermi 5, Firenze, I-50125, Italy
             \and 
             Dipartimento di Fisica e Astronomia, Universitá degli Studi di Firenze, via G. Sansone 1, 50019, Sesto Fiorentino, Italy
             \and
             Max-Planck-Institut für Astrophysik, Karl-Schwarzschild-Str. 1, D-85748 Garching, Germany
             \and
             Dipartimento di Fisica “G. Occhialini”, Università degli Studi di Milano-Bicocca, Piazza della Scienza 3, 20126 Milano, Italy
             \and
             Scuola Normale Superiore, Piazza dei Cavalieri 7, I-56126 Pisa, Italy
            \and
            NASA Goddard Space Flight Center, Greenbelt, MD 20771, USA
            \and
            INAF – Osservatorio di Astrofisica e Scienza dello Spazio di Bologna, via Gobetti 93/3, I-40129, Bologna, Italy
            \and
            INAF - Osservatorio Astronomico di Trieste, Via G. B. Tiepolo 11, I–34131 Trieste, Italy
            \and 
            IFPU - Institut for fundamental physics of the Universe, Via Beirut 2, 34014 Trieste, Italy
            \and
            Kavli Institute for Cosmology, University of Cambridge, Madingley Road, Cambridge, CB3 0HA, UK
            \and
            Cavendish Laboratory Astrophysics Group, University of Cambridge, 19 JJ Thomson Avenue, Cambridge, CB3 0HE, UK
            \and 
            Department of Physics and Astronomy, University College London, Gower Street, London WC1E 6BT, UK
            \and
            Instituto de Estudios Astrofísicos, Facultad de Ingeniería y Ciencias, Universidad Diego Portales, Av. Ejército Libertador 441, Santiago, Chile
            \and
            Kavli Institute for Astronomy and Astrophysics, Peking University, Beijing 100871, People’s Republic of China
            \and
            Dipartimento di Fisica, Università di Roma La Sapienza, Piazzale Aldo Moro 2, I-00185 Roma, Italy
            \and
             INFN – Sezione Roma1, Dipartimento di Fisica, Università di Roma La Sapienza, Piazzale Aldo Moro 2, I-00185 Roma, Italy
             \and
             Sapienza School for Advanced Studies, Viale Regina Elena 291, I-00161 Roma, Italy
             \and
              Dipartimento di Fisica e Astronomia ‘Augusto Righi’, Università degli Studi di Bologna, Via P. Gobetti, 93/2, 40129 Bologna, Italy
             }
\authorrunning{I. Villani et al.}

   \date{}

 
  \abstract
   {
   The most luminous and obscured quasars (QSOs) detected through sensitive infrared all-sky surveys are believed to represent a key co-evolutionary phase from nuclear to circum-galactic (CG) scales in the formation of massive galaxies. In this context, Hot Dust Obscured Galaxies (Hot DOGs) in the redshift interval z $\sim$ 2\,$-$\,4 (the so-called Cosmic Noon) provide unique opportunities to investigate the relationship between cosmic mass assembly and the nuclear accretion processes of luminous QSOs/galaxies at high-z. W0410\,$-$\,0913 (hereafter W0410\,$-$\,09) is a luminous ($\rm L_{bol} \sim 6.4\times 10^{47}\, \rm erg \,s^{-1}$) and obscured QSO at z = 3.631, characterized by a 30 kpc CG Ly$\alpha$ nebula (CGLAN), rather small if the $\sim$ 100 kpc Ly$\alpha$ nebulae around the unobscured QSO compared to the Type-I QSO peers, and by an exceptional overdense environment of Ly$\alpha$ emitters (LAEs) with $\sim$ 19 of them located in the CG region of 300 kpc at $\pm\, 200 \,\rm km \,s^{-1}$ from the Hot DOG.}
   {Our aim is to detect and characterize active nuclear accretion in the Hot DOG W0410\,$-$\,09 and its environment.}
   {We carry out this study by exploiting a deep proprietary
    $\sim$ 280 ks Chandra X-ray Observatory observation. We employ a set of empirical models suited for obscured sources and physically motivated spectroscopic models to account for a toroidal X-ray obscurer and the reprocessing of the X-ray radiation.}
  {We find that the W0410\,$-$\,09 consistently exhibits nuclear obscuration levels from mild to heavy star formation (Compton-thick, CT), with hydrogen column density of $\rm N_{H} > 10^{24} \,\rm cm^{-2}$ (and up to $\rm N_{H} \sim 10^{25}\, cm^{-2}$) and a intrinsic luminosity of $\rm L_{2-10} > 10^{45}\, \rm erg\,s^{-1}$. W0410\,$-$\,09 is therefore one of the most luminous and obscured QSO at z\,$>$\,3.5 discovered so far. This level of obscuration and the highly accreting nature of the Hot DOGs suggest that W0410\,$-$\,09 is undergoing a blow-out phase. This phase is predicted by models of merger-driven QSO formation scenarios, where strong winds begin to clear the dusty obscuring medium from the nuclear surroundings. We speculate that this heavy nuclear obscuration limits the amount of UV disc emission powering its CGLAN, thereby likely explaining its small nebula size. With the exclusion of W0410\,$-$\,09 we do not detect X-ray emission from any of the 19 LAEs. We analyze their combined emission in several bands finding significant signal at the 3$\sigma$ level only in the 6\,$-$\,7 keV rest-frame energy band which we interpret as due to Fe K$\alpha$ line. This strongly suggests the presence of heavily obscured yet undetected AGN emission in several LAEs. Considering W0410\,$-$\,09 estimate an AGN fraction of $f_{\mathrm{AGN}}^{\mathrm{LAE}} =5^{+12}_{-4} \%$. This value can be as high as $\sim35\%$ if we account for the presence of unresolved obscured AGN as suggested by the Fe K$\alpha$ line detection.

 }
   {W0410\,$-$\,09 is powered by an intrinsically luminous, CT quasar. Its high obscuration likely explains the limited extent of its CGLAN. Our analysis suggests that this object is in a crucial transitional blow-out phase, during which powerful QSO-driven outflows will sweep out the nuclear obscuration, paving the way for an unobscured bright quasar.
   }
   {}

   \keywords{X-rays: galaxies -- galaxies: active -- quasars: general -- quasars: individual: WISE 0410\,$-$\,0913
               }

   \maketitle
%

\section{Introduction}
Hot Dust-Obscured Galaxies (Hot DOGs) represent a rare population of hyper-luminous galaxies \citep[$\sim$ 1000 known across the entire sky;][]{wu2012}, discovered in the WISE all-sky survey using the 'W1W2-dropout' method. These galaxies are significantly detected in the WISE 12 $\rm \mu m$ (W3) and 22 $\rm \mu m$ (W4) bands but are faint or undetectable in the 3.4$\rm ,\mu m$ (W1) and 4.6$\rm ,\mu m$ (W2) bands. 
Their spectral energy distribution (SED) peaks in the mid-infrared, driven by dust emission at temperatures of around 60\,$-$\,100 K \citep[e.g.,][]{fan2016b}, significantly warmer than the 30\,$-$\,40 K typically found in normal MIR-selected sources, including dust-obscured galaxies \citep[][]{dey2008,magnelli2012}. This indicates that these sources are powered by Active Galactic Nuclei (AGN) related processes. Previous research shows that Hot DOGs are exceptionally luminous \citep[$\rm L_{bol}\!>\!10^{13} \,L_{\odot}$;][] {tsai2015}, heavily dust-obscured \citep{Eisenhardt2012} quasars (QSOs) at high redshift ($\rm z \geq 2$), which are believed to represents a unique and short-lived phase between starburst-dominated and optically bright QSOs in merger-driven QSO formation scenarios \citep{sanders1988,hopkins2006}. Indeed, these models suggest that galaxy mergers induce gas inflow toward the host and nuclear regions, triggering intense star formation and massive nuclear SMBH accretion. These high mass accretion,
responsible for the bright AGN activity, could at the same time
be responsible for heavy nuclear obscuration, which is efficiently
removed, on relatively short time-scales during the so-called
blow-out phase, by strong AGN-driven outflows leading, at the
end, to an unobscured luminous blue QSO. Multiwavelength observations corroborate the interpretation of the Hot DOGs in this scenario by reporting high SFR \citep{fan2016b}, presence of nuclear and host-wide winds \citep{fan2018, ginolfi2022} and association with dense protocluster-like environments \citep{jones2014,diazsantos2018,ginolfi2022}. In particular, the very high X-ray luminosities and presence of heavy nuclear obscuration up to star formation levels (CT, i.e. $\rm N_{H}\geq 1.5 \times 10^{24} \,\rm cm^{-2}$) as revealed by X-ray spectroscopic analysis \citep{piconcelli2015,ricci2017,zappacosta2018, vito2018}, place these sources in the transitional obscured stage. Multiwavelength observations further suggest that Hot DOGs are  found in dense galactic environments indicative of protocluster regions \citep{jones2014, ginolfi2022, zewdie2025} and are associated with galaxy mergers \citep{fan2018,diazsantos2018}. As such, Hot DOGs provide at all scales valuable insights into QSO/host co-evolution and SMBH growth during intense AGN phases. 

In this work, we present the Chandra observation of the Hot DOG W0410\,$-$\,09 (RA: 4:10:10.640 deg, DEC: -9:13:05.380 deg, epoch: J2000) at $\rm z\!=\!3.631$ \citep{diaz2021,stanley2021}. This source is one of the brightest \citep[$\rm L_{bol} \sim 6.4\times 10^{47}\, \rm erg \,s^{-1}$, $\rm L_{IR} \sim 2 \times 10^{14} \,L_{\sun}$,][]{fan2016,diaz2021}, and most massive and gas-rich systems identified to date, with stellar ($\rm M_{star}$) and molecular gas ($\rm M_{gas}$) masses of $>10^{11}\rm M_{\odot}$ \citep{fan2018,diaz2021}. It is also characterized by a high star formation rate $>1000 \,\rm M_{\odot} yr^{-1}$ \citep{frey2016}. The QSO and its circum-galactic (CG) UV emission are studied by \citet{ginolfi2022} using the MUSE integral field spectrograph at VLT. The MUSE spectrum reveals broad (FWHM $\sim 2800$ km s$^{-1}$), blueshifted NV $\rm \lambda 1240$ and {CIV $ \rm \lambda\lambda 1548, 1550$ emission lines, hinting to the presence of AGN-driven nuclear outflows. Additionally, \citet{ginolfi2022} detect a narrow (FWHM $\sim 400$ km s$^{-1}$) Ly$\alpha$ line blueshifted by about 1400 km s$^{-1}$. 
Integral field and narrow band spectroscopic observations indicate that virtually all luminous unobscured QSOs are surrounded by giant CG Ly$\alpha$ nebulae (CGLANs), vast cosmic structures reaching 100 kpc in extent, with gas temperatures around $10^{4}$ K \citep[]{cantalupo2014,martin2014, hennawi2015, arrigoni2018}. While giant CGLANs are commonly observed around Type-I QSOs, their presence in the vicinity of Type-II QSOs like Hot DOGs remains poorly explored \citep{bridge2013}.
 For W0410\,$-$\,09 object the CG diffuse Ly$\alpha$ emission is relatively modest with an extension of $\sim 30 \,\rm kpc$. Furthermore, the target is located in a unique highly overdense CG region with $>19$ Ly$\alpha$ emitting companions, connected by Ly$\alpha$ filamentary structures over a 300 kpc scale \citep{ginolfi2022}. This relatively weak Ly$\alpha$ emission, despite the extreme luminosity of the source, could be due to the heavy obscuration of the AGN, which likely prevents the ionizing UV flux from powering the nebula out to large scales. Such a scenario is supported by recent findings for sub-millimeter-bright QSOs, where the lack of extended Ly$\alpha$ emission has also been attributed to strong nuclear obscuration \citep{gonzalezlobos2023}. 
 
This paper analyzes a $\sim\!280 \rm \,ks$ Chandra observation of W0410\,$-$\,09, aiming to (i) test the hypothesis that high obscuration is responsible for the compact CGLAN and (ii) investigate the occurrence of AGN in the W0410\,$-$\,09 Ly$\alpha$ companions in such an overdense environment.
 
 The paper is organized as follows: Sect. \ref{reduction} outlines the Chandra data reduction steps. Sect. \ref{analysis} presents an extensive X-ray spectral and photometric analysis of W0410\,$-$\,09, while Sect. \ref{X-ray emission companions} focuses on the X-ray emission from its companion sources. Sect. \ref{discussion} is devoted to discussion, and our conclusions are presented in Sect. \ref{conclusions}. Standard astronomical orientation N up E left is adopted. Throughout the paper we assume a cosmology with $\Omega_\Lambda = 0.73$ and $\rm H_0 = 70\,\mathrm{km\,s^{-1}\,Mpc^{-1}}$. Henceforth errors correspond to the 68.3\% (1$\sigma$) confidence level for one interesting parameter, while the lower limits are reported at 90\% confidence level.

\section{Data reduction}\label{reduction}
The data for the source analyzed in this study are obtained from the WebChaser archive hosted on the Chandra X-ray Center website\footnote{https://cxc.harvard.edu/}. The target was observed for $\sim$ 280 ks (P.I. L. Zappacosta) in 12 different exposures, obtained from 13 November 2021 to 13 November 2022, ranging from 15 to 33 ks.
Details of each exposure are reported in Table \ref{table:1}.


\setlength{\tabcolsep}{3.0pt}
\begin{table}[h]
\begin{center}
\small
\caption{Details of W0410\,$-$\,09 ACIS-S Chandra observations.}
\label{table:1}
\begin{tabular}{ccc@{\hspace{1cm}}ccc}
\hline\hline\\
ID & Exp & Date & ID & Exp & Date \\
       & (ks) &      &        & (ks) &      \\
\hline\\
25836 & 23 & 2021-11-13 & 25837 & 25 & 2021-11-28 \\
25832 & 26 & 2021-11-16 & 25834 & 15 & 2021-12-09 \\
25383 & 28 & 2021-11-21 & 26224 & 15 & 2021-12-09 \\
25830 & 30 & 2021-11-23 & 25829 & 28 & 2022-11-12 \\
25833 & 33 & 2021-11-24 & 25831 & 23 & 2022-11-13 \\
25835 & 20 & 2021-11-27 & 27543 & 16 & 2022-11-13 \\ \\
\hline
\end{tabular}
\end{center}
\footnotesize{
\noindent The first column presents the Chandra Observation Identification Number (ID), followed by the second column, which specifies the actual exposure time of the observation in kiloseconds (ks). The third column displays the date of the observation.
}
\end{table}

 Then, the \textit{merge\_obs} script is used to obtain a sum of the event files from the different exposures. This script processes a stack of event files by re-projecting them into a shared tangent point, merging them into a cohesive dataset, generating exposure maps for individual observations, and ultimately dividing the resultant images to yield a coadded, exposure-corrected image. Due to the small number of sources detected in the field, it is not possible to perform a reliable absolute astrometric correction by matching to external catalogs. Hence, the subsequent data analysis is performed on the uncorrected images. However, the typical Chandra astrometric accuracy is less than 0.4 arcsec and we estimate in $\sim$ 0.2\,$-$\,0.3 arcsec the relative astrometric alignment, using the longest observation as reference (i.e. 25833).
 Three images are generated from the new event file, within specific energy ranges: $0.3\!-\!7 \rm \,keV$ (full band, see Fig. \ref{fig1}), $0.3\!-\!2\,\rm \,keV$ (soft band) and $2\!-\!7 \rm \,keV$ (hard band). These images are created for the back-illuminated ACIS-S3 chip containing the observed target at the aimpoint of the instrument. To derive the X-ray coordinates of the source and identify any potential contaminant sources around the target, we employ the source detection algorithm \textit{wavdetect}, running it separately in the three energy bands: soft, hard and broad. The source is well detected at a position of RA: 4:10:10.6538 and DEC: -9:13:05.893 (J2000), which differs by 0.55~arcsec from the optical position reported in \citet{ginolfi2022}.
We choose a circular extraction region of radius 2 arcsec centered on the X-ray position of the target and containing $\gtrsim$ 95\% of the source to perform photometry and spectral extraction (Fig. \ref{fig1}, left panel). For the background we use a circular region, centered on the target, of radius 60 arcsec. From this region, we remove the point sources detected through the \textit{wavdetect} tool by adopting a circular aperture of radius 3~arcsec except for W0410\,$-$\,09, the brightest source in this region, for which we adopt an aperture of 6 arcsec radius (Fig. \ref{fig1}, right panel). We extract source and background counts from these regions in each energy band. In the three bands we obtain ${74.0}_{-8.7}^{+9.3}$ (full band), ${20.8}_{-4.6}^{+5.2}$ (soft band) and ${53.2}_{-7.2}^{+7.9}$ (hard band) background-subtracted counts. Net-counts and $1\sigma$ uncertainties are estimated by calculating the no-source binomial probability \citep[][]{weisskopf2007,vito2019}.
\\We perform the spectral extraction using the \textit{specextract} CIAO script which creates source and background spectra and corresponding response files. We combine all the spectra and response files using the \textit{addascaspec}
script from the FTOOLS X-ray data and FITS files manipulation package\footnote{https://heasarc.gsfc.nasa.gov/ftools/} to produce summed source and background spectra and correspondingly combining the response
files. 

\begin{figure}[h!]
 \centering
{\includegraphics[width=0.49\textwidth] {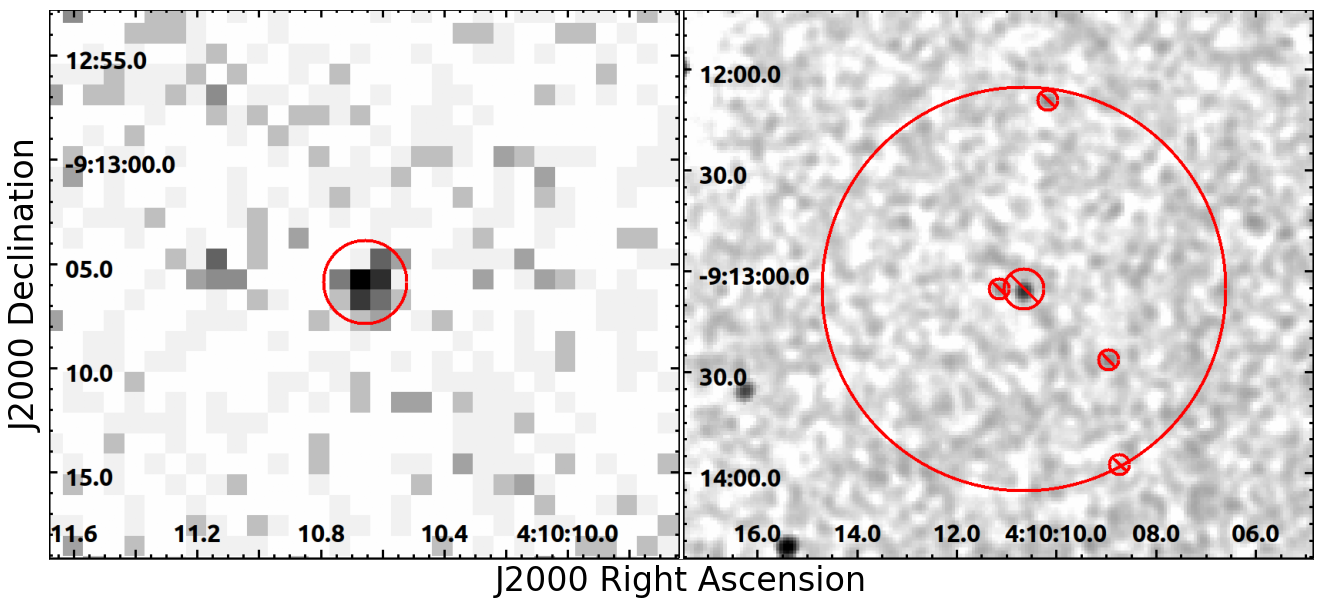}}
\caption{Left panel: ACIS-S image of W0410\,$-$\,09 in the 0.3\,$-$\,7 keV energy band. The red circle shows the 2 arcsec wide region used. Right panel: the circular region used from the extraction of the background. Sources detected by \textit{wavedetect} and W0410\,$-$\,09, and reported as crossed out circular
regions, are not considered.}
\label{fig1}
\end{figure}

\section{Spectral analysis of W0410\,$-$\,09}\label{analysis}
The XSPEC (version 12.12.1) spectral fitting software is used for spectral analysis. All the combined spectra are binned using the optimal method described by \citet{kaastra2016} and analyzed using the Cash statistic implemented in XSPEC with direct background subtraction \citep[W-stat;][]{cash1979,wachter}. A Galactic column density of $\rm N_{H}^{Gal} = 4.03 \times 10^{20} \,\rm cm^{-2}$ is adopted \citep{floer}. The spectral analysis is performed in the 0.3\,$-$\,7 keV energy range. 
\subsection{Empirical models}\label{empirici}

We start with a power-law model (\textsc{Pow} model) modified only by the absorption from our Galaxy interstellar medium, which we parametrized with the \textsc{tbabs} model. This results in a best-fit model (C-stat/d.o.f. = 46.1/48, Fig. \ref{pow}) with a photon index $\rm \Gamma\!=\! 0.20 \pm{0.25}$. This value is much flatter than the typical value of $\rm \Gamma\!\sim\!$ 1.8\,$-$\,1.9 reported for QSOs \citep{piconcelli2005, degliagosti2025} and strongly suggests that the source is highly obscured. This model indeed results into an X-ray bolometric correction $\rm K_{bol,x} = L_{bol}/L_{2-10} \!\sim\!6600$, a value that appears unreasonably large compared to expectations \citep{Duras2020}. Accordingly, the \textsc{Pow} model is modified by an intrinsic absorption term (i.e. at the redshift of W0410\,$-$\,09) parametrized by the \textsc{ztbabs} multiplicative model. In this case, assuming $\Gamma=1.9$, a very high value of $\rm N_{H}$ is obtained ($\sim10^{24}~\rm cm^{-2}$). For very high absorption column density, it is necessary to also consider the effect of Compton scattering. Hence, the absorbed model is further modified with the \textsc{cabs} model \citep[][XSPEC User’s Guide v12.12.1]{arnaud1996} to account for Compton scattering of X-ray photons. This model only takes into account the loss of photons outside of the line of sight. Adopting this model (\textsc{CabsPow}) we obtain a best-fit parametrization (C-stat/d.o.f.$ = 54.8/48$, Fig. \ref{pow}) with $\rm \nh={0.9}_{-0.2}^{+0.2} \times 10^{24} \rm cm^{-2}$.  Positive residuals at the position of the Fe K-lines (6\,$-$\,8~keV) are clearly visible. Given the existence of Al K$\alpha$ and Si K$\alpha$ background lines, at observed energies between 1.5\,$-$\,2~keV (consistent with the rest-frame energies of the positive residuals), we check if these could be due to an improper background subtraction. Hence, we try different background circular regions at distances up to $\sim$ 1.8 arcmin from the Hot DOG placed toward different directions. We conclude that the residuals are not due to a background artifact. A Gaussian line left free to vary in this energy range results in a best-fit energy of 7.9 $\pm \,0.2$~keV (C-stat/d.o.f. = 50.5/46). This component, consistent with Fe XXV K$\beta$ line, does not significantly improve the fit ($\rm P_{null}\!=\!0.15$, according to an F-test). We further investigated if the use or addition of further lines (i.e. Fe K$\alpha$ at 6.4~keV and Fe XXV K$\alpha$ line at 6.697~keV) could improve the modeling.
However the fits do not improve considerably and the values of the parameters obtained with the model including the lines do not change significantly. Notice that for Fe K$\alpha$ line we obtain a rest-frame equivalent width upper limit of 1.6 keV. A high equivalent width Fe K$\alpha$ line is commonly associated with reflection in CT AGN where the primary emission is entirely suppressed as typically reported in CT AGN \citep[e.g.,][]{Ghisellini1994, Matt2003, Risaliti2004, hickox2018}. Therefore, the Fe K$\alpha$ upper limit is entirely consistent with the extreme CT obscuration estimated by this model.
Hence, the line components were not included in the final best-fit \textsc{CabsPow} model.

Given the high N$\rm_{H}$, we also evaluate the possibility that the primary power-law spectrum is completely absorbed by matter with \mbox{$\rm N_{H} \gg 10^{24} cm^{-2}$} and therefore completely dominated by the reflection component from cold material parameterized by a \textsc{pexmon} model (see Fig. \ref{pow}). The reflection model employs a planar geometry with infinite optical depth, illuminated by the primary continuum. For an isotropic source, it spans an opening angle of \mbox{$\Omega$ = 2$\pi$ $\times$ R}, where R is a parameter representing the reflection strength. We adopt this model under certain assumptions: solar abundance, an exponential high-energy cut-off (\mbox {$\rm E_{c} = 200~keV$}) applied to the primary power-law incident radiation \citep[e.g.,][]{fabian} and a reflector inclination angle of 60 deg. In this model, only the reflection component is taken into account, neglecting the primary one that for our purposes is considered completely absorbed (model \textsc{ReflDom}). The best-fit parametrization provides an equally good but simpler fit to the data, with only the normalization free to vary (C-stat/d.o.f. = 48.8/49). This suggests that the primary coronal component is obscured by a heavy CT absorber which provides
an equally good statistical representation, compared to the previous fit of the Hot DOG spectrum. The luminosity in the \mbox{2\,$-$\,10~keV} range of the reflected component is L$_{2-10} = (1.0 \pm 0.1) \times 10^{44} \rm~erg~s^{-1}$. Since the N$\rm_{H}$ of the absorber is unconstrained in this case, it is not possible to calculate the absorption-corrected (i.e. intrinsic) luminosity of the coronal component. The derived parameters for all the empirical models used in our analysis are reported in Table \ref{table:empiricalmodel}.

\begin{figure}[h!]
  \centering
  \includegraphics[width=0.49\textwidth]{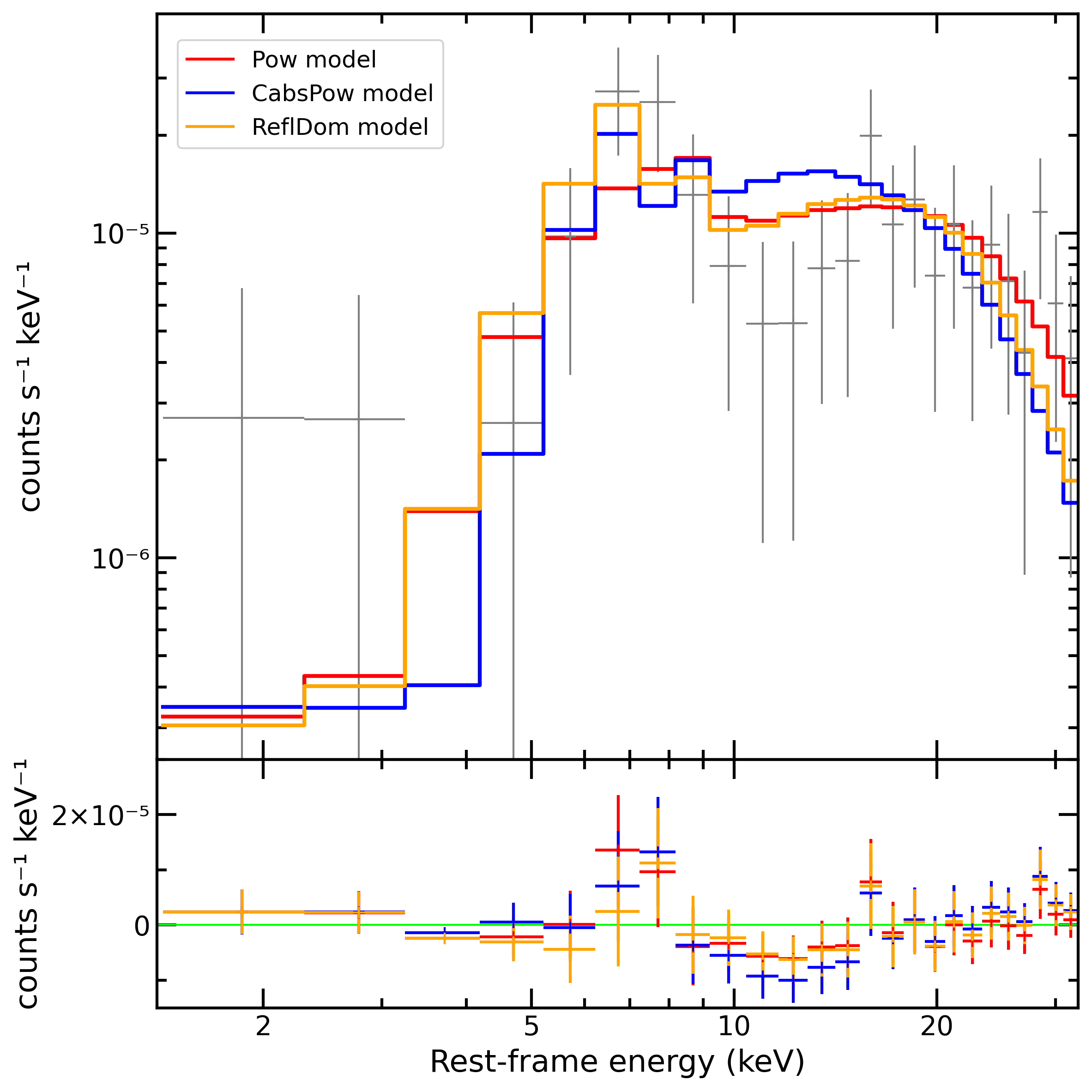}
  \caption{Chandra/ACIS-S spectrum of W0410\,$-$\,09. Empirical models are reported in different colors: \textsc{Pow} model in red, \textsc{CabsPow} model in blue and \textsc{ReflDom} model in orange. The spectrum has been slightly rebinned for better visualization. In the lower panel, the residuals for each model are shown in the same color as their respective model.}
  \label{pow}
\end{figure}

\subsection{Geometry-dependent obscuration models}
 To obtain more accurate and physically motivated constraints, we use two different 
models derived from Monte Carlo radiative transfer simulations employing a toroidal geometry for the obscurer and in which the X-ray source is located at the geometrical center of the absorbing structure which obscures and reprocesses the X-ray power-law shaped emission. The first model approximates the torus geometry by employing a spherical absorber, with polar cut-outs with variable half-opening angle $\rm \theta_{tor}$ \citep[][hereafter  \textsc{Borus}] {balokovic}. The second model employs a doughnut torus geometry with a fixed $ \rm \theta_{tor}$ = 60 deg \citep[][hereafter \textsc{MYTorus}] {Murphy,yaqoob}.
Both models adopt a uniform density obscurer, where the gas is assumed cold and neutral and takes into account the Compton scattering. Fluorescent line emission, and the reprocessed continuum, are calculated self-consistently. The gas is assumed to be uniformly distributed, with elemental abundances similar to that of the Sun. 
Neutral K-shell, $\alpha$ and $\beta$ transitions at $\sim$\,6\,$-$\,7 keV are calculated for Fe and Ni by both models. \textsc{Borus} includes K$\alpha$ and K$\beta$ transitions from elements up to zinc ($Z<31$). 

\begin{table*}[t]
\begin{center}
\caption{Best-fit parameters derived from the empirical models.}
\label{table:empiricalmodel}
\begin{tabular}{lcccccc} \hline\hline \\
Model & $\Gamma$ & C-stat/d.o.f. & $\rm N_{H}$ & Flux$_{0.5-2}$& Flux$_{2-10}$ &  L$_{2-10}$\\ 
& & & {\tiny(10$^{24}$\textit{\rm cm$^{-2}$})} & {\tiny(10$^{-16}$\textit{\rm erg cm$^{-2}$ s$^{-1}$})} & {\tiny(10$^{-15}$\textit{\rm erg cm$^{-2}$ s$^{-1}$})} & {\tiny(10$^{45}$\textit{\rm erg s$^{-1}$})} \\ \hline\\
\textsc{Pow$^{(a)}$} &  ${0.2}_{-0.3}^{+0.3}$ & 46.1/48 & - & ${5.4}_{-2.4}^{+0.5}$ & ${10.9}_{-4.8}^{+0.7}$ & ${0.09}_{-0.04}^{+0.08}$ \\
\textsc{CabsPow$^{(b)}$}& (1.9) & 54.8/48 & ${0.9}_{-0.2}^{+0.2}$ &${4.3}_{-1.0}^{+0.9}$ & ${6.2}_{-1.0}^{+0.8}$& ${1.7}_{-0.4}^{+0.6}$\\
\textsc{ReflDom$^{(c)}$} & (1.9) & 48.8/49 & - &${6.3}_{-0.7}^{+0.8}$ & ${6.5}_{-0.8}^{+0.8}$& ${0.10}_{-0.01}^{+0.01}$$^{(d)}$\\\\ \hline\hline
\end{tabular}
\end{center}
{\footnotesize The parameters in brackets are held constant at the specified value. $^{(a)}$\textsc{tbabs*zpow}; $^{(b)}$\textsc{tbabs*ztbabs*cabs*zpow};
$^{(c)}$\textsc{tbabs*pexmon};
$^{(d)}$ Luminosity of the reflected component.}
\label{table:2}
\end{table*}

\subsubsection{Borus model}
\textsc{Borus} assumes a primary spectral component which has the form of a power-law with an  high-energy exponential cut-off and provides a table model including only the reprocessed spectral components (both continuum and lines). Hence it needs to be used in conjunction with the primary component. We fit the ACIS spectrum adopting this model in XSPEC: 
\begin{equation}
\textsc{tbabs}~(\textsc{ztbabs} \times \textsc{cabs} \times \textsc{zcutoffpl} + \textsc{Borus}+[\textsc{zgauss}])
\end{equation}

where \textsc{tbabs} is the photoelectric Galactic absorption term, \textsc{zcutoffpl} is the primary redshifted power-law component with an high-energy exponential cut-off, which is modified by the photoelectric absorption at the redshift of the source (\textsc{ztbabs}), and the Compton scattering terms (\textsc{cabs}). \textsc{Borus} is the torus table model which accounts for the reprocessing by the geometrical toroidal structure.
In our fits we will also evaluate the need for an additional Gaussian component (\textsc{zgauss}) to better describe the possible presence of an ionized Fe K-line at $\sim$ 7\,$-$\,8 keV rest-frame (see Sect. \ref{empirici}).

 The \textsc{Borus} table model only accounts for $\Gamma$ values in the range 1.4\,$-$\,2.6. Given the small number of spectral counts, $\Gamma$ would be loosely constrained over this large range. Therefore, we will perform our fits with $\Gamma$ fixed to 1.9, which is the slope typically expected for AGN and measured in QSOs at Cosmic Noon and in this luminosity regime \citep[e.g.,][]{nardini2019, zappacosta20}.
We also set $\rm \theta_{tor}$, inclination angle ($\rm \theta_{inc}$), cut-off energy ($\rm E_{cut}$) and iron abundance ($\rm A_{fe}$) to 60~deg (equal to \textsc{MYTorus} for a consistent comparison), 80~deg (i.e. almost edge-on), 200~keV and Solar, respectively. The best-fit \textsc{Borus} model gives $\rm N_H \sim 0.9 \times 10^{24}\, \rm cm^{-2}$ (Cstat/d.o.f. = 54.6/48, see Fig. \ref{borus01} left panel). We include a \textsc{zgauss} component to account for residual at $\sim$ 7\,$-$\,8 keV rest-frame. The best-fit modeling, resulting in line energy of $\rm E_{gauss}=7.9\pm{0.2}~$ keV (C-stat/d.o.f. = 48.4/46), does not significant improve the fit ($\rm P_{null}=0.06$ according to an F-test). We try also a model with $\rm \theta_{tor}=0$ (i.e. a sphere with no polar cut-outs; hereafter \textsc{BorSphere}). In this case,
 the best-fit model does not need a \textsc{zgauss} component and returns a value of $\rm N_{H} \sim 1.3 \times 10^{24}\, cm^{-2}$, but with a worse fit statistics (C-stat/d.o.f. = 58.3/48), exhibiting large residuals (see Fig. \ref{borus01} right panel) across the entire energy range. The derived parameters for  \textsc{Borus} and \textsc{BorSphere} modeling are reported in Table \ref{table:3}. 

\begin{figure*}[h!]
 \centering
{\includegraphics[width=0.5\textwidth] {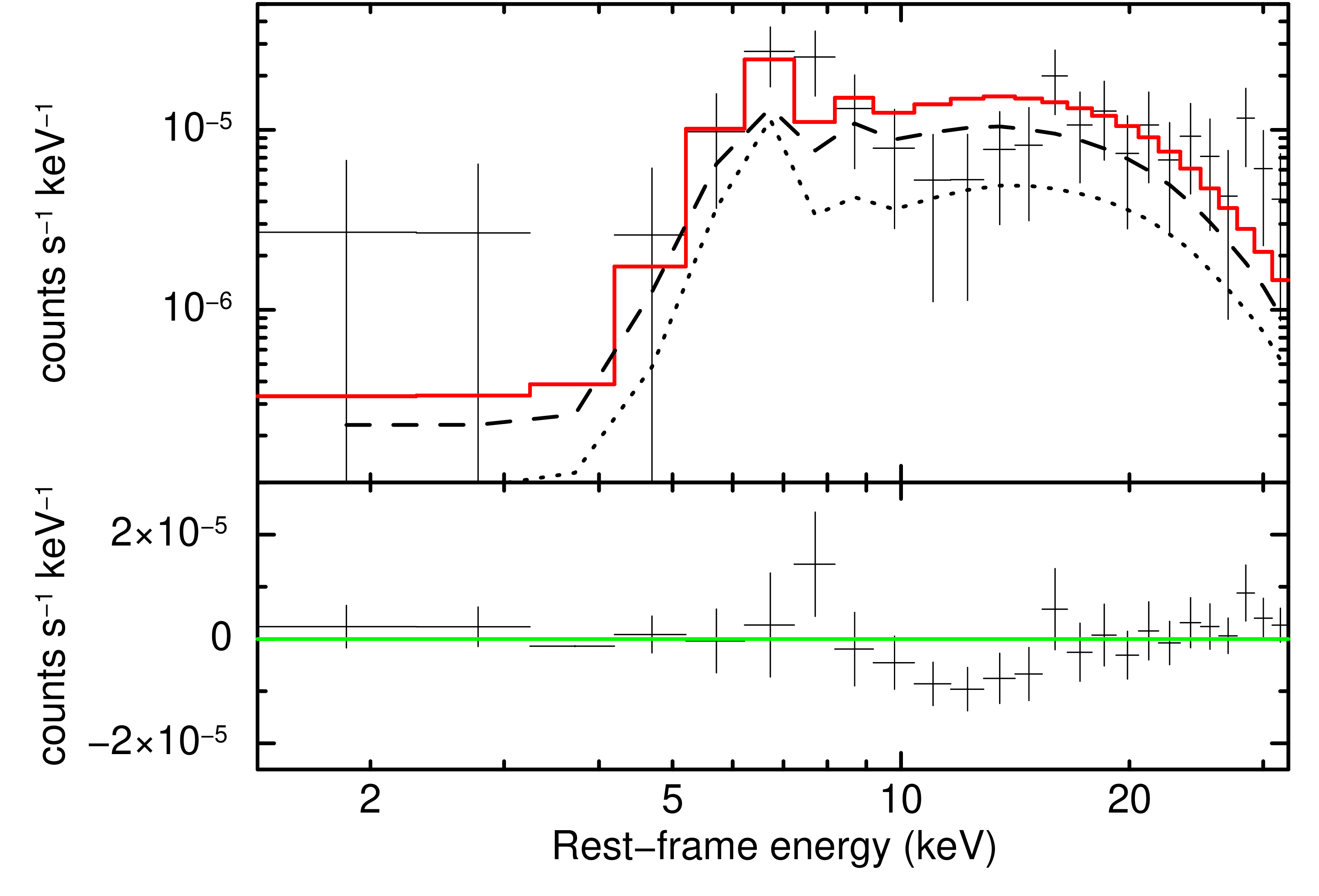}}{\includegraphics[width=0.5\textwidth] {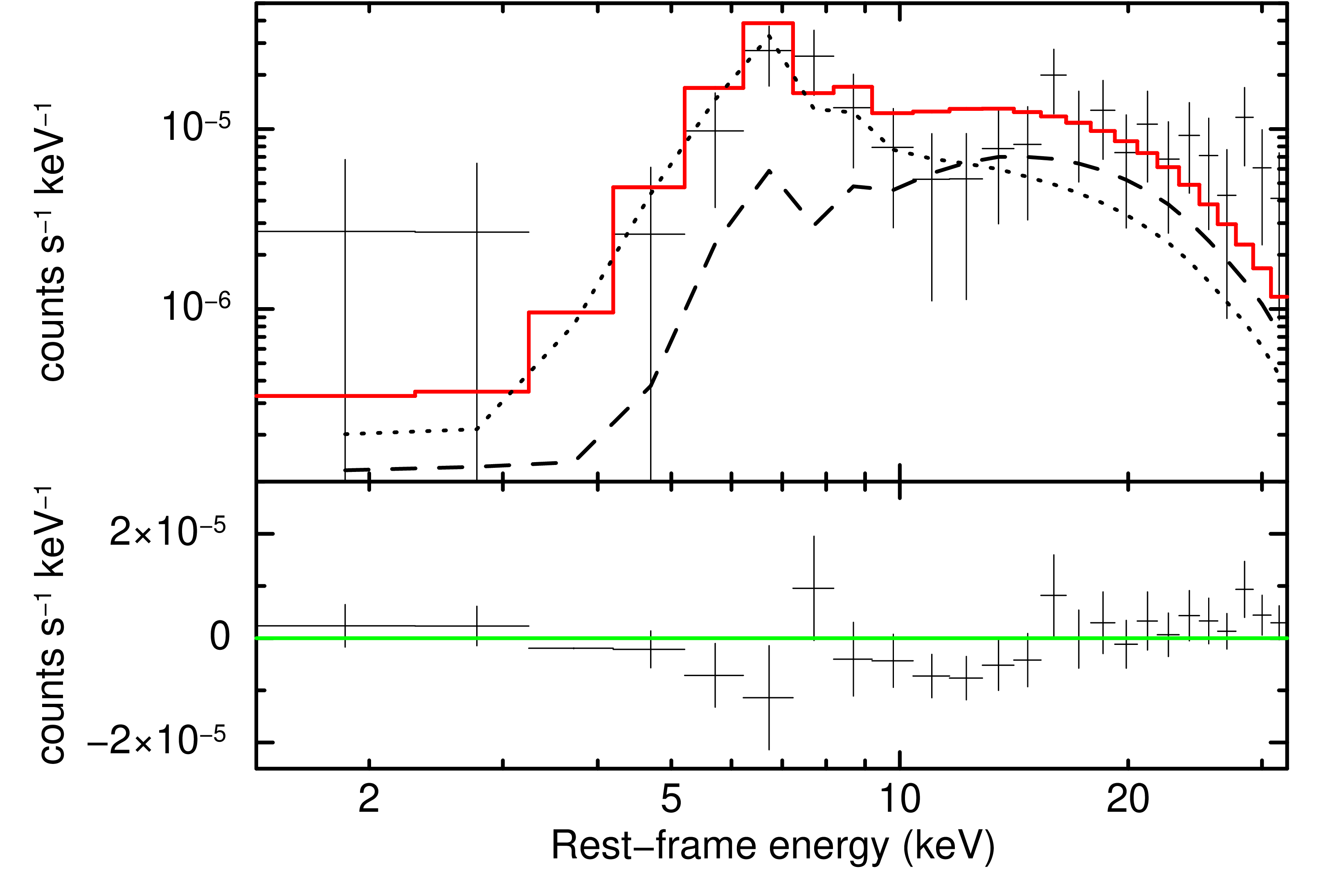}}
\caption{Left panel: Chandra/ACIS-S spectrum with \textsc{Borus} model and residuals. The red solid line represents
the best-fit model, the dotted line represents the reflection component and the dashed line represents the heavily-absorbed power-law continuum with exponential cut-off. The spectra have been slightly rebinned for better visualization.
 Right panel: Chandra/ACIS-S spectrum with \textsc{BorSphere} model and residuals. The red solid line represents the best-fit model, the dotted line represents the Compton reflection component, the dashed line indicates the absorbed cut-off power-law component.}
\label{borus01}
\end{figure*}

 \subsubsection{MYTorus model}
 The \textsc{MYTorus} implementation in XSPEC consists of three different table model components: (1) one for the attenuation of the line of sight radiation due to photoelectric and Compton-scattering effects (MYTZ); (2) one to reprocess the Compton-scattered radiation (MYTS); (3) and one that calculates the contribution from fluorescent line emission (MYTL).
 Therefore the XSPEC parametrization of this model is:

\begin{equation}
\textsc{tbabs}~( \textsc{zpow} \times \textsc{MYTZ} + c_{s} \times \textsc{MYTS} + c_{l} \times \textsc{MYTL} +[\textsc{zgauss}])
\end{equation}

where \textsc{zpow} is the redshifted power-law component, c$_s$ and c$_l$ are the normalization constants of MYTS and MYTL, respectively. 
 In our parametrization the three components are applied progressively in sequence, adding one at a time to the previous one to determine the best model description to the data. The three components initially share the same absorber column density and are combined with normalization constants (c$_s$ and c$_l$) initially set to unity. To match the geometric requirements imposed by standard unification schemes where Type-II sources are observed at high inclinations and to account for the high column density, we set an almost edge-on view of the torus at $\rm \theta_{inc} = 80 \rm~deg$. We assume $\Gamma$=1.9 for the slope of the continuum power-law. \textsc{MYTorus} provides an estimate of the equatorial column density, $\rm {N}_{H}^{eq}$, which is defined as the equivalent hydrogen column density through the diameter of the tube of the torus. The
actual line of sight $\rm N_{H}$, which is the $\rm N_{H}$ quantity measured in the empirical models and in \textsc{Borus} for this particular torus aperture ($\rm \theta_{tor} = 60 \rm~deg$) translates into $\rm N_{H} \approx 0.94~ \rm N_{H}^{eq}$.
All components are connected to the same $\rm {N}_{H}^{eq}$.

We start by employing a model using the MYTZ component. We obtain a best-fit model with C-stat/d.o.f. = 54.4/48 and a column density $\rm N_{H} \sim 0.9 \times 10^{24} \,\rm cm^{-2}$. 
The poor parametrization is mainly given by the large negative residuals at the energies 10\,$-$\,15 keV rest-frame, corresponding to 2\,$-$\,3 keV observed energies. Therefore we add a scattered \textsc{MYTS} component. This results in an $\rm N_{H}\!\sim 0.9 \times 10^{24} \,\rm cm^{-2}$ is obtained with a \mbox{C-stat/d.o.f. = 54.2/48}. We also attempt to decouple the $\rm N_{H}$ values of the two components, considering that the absorbed portion along the line of sight may have a different column density than the scattered medium. This test considerably improve our fit with C-stat/d.o.f. = 46.5/47 but requiring a line of sight column density of $\rm N_{H}\!\sim 4.6 \times 10^{24} \, cm^{-2}$ and column density from the \textsc{MYTS} component of only $10^{22} \rm cm^{-2}$ which is highly unlikely and would point to heavy obscuration exclusively along the line of sight. Hence we re-couple the $\rm N_H$ values and add a line component (MYTL) to account to line residuals at 6\,$-$\,7 keV. We obtain a best-fit column density of $\rm N_{H}\!\sim 1.0 \times 10^{24}\,\rm cm^{-2}$ (C-stat/d.o.f.= 53.7/48). The model shows a positive residual at 7\,$-$\,8 keV. Following the previous attempts we add and additional Gaussian component (\textsc{zgauss}) to account for the positive residual at $\sim$ 7.9 keV obtaining a best-fit model with C-stat/d.o.f. = 47.1/46. An F-test between the models with and without the \textsc{zgauss} component returns a $\rm P_{null}$ = 0.05, indicating a marginal statistical improvement, however the rest-frame equivalent width is 3.6 keV which is extreme for such lines in heavily obscured AGN. Therefore, we prefer to ascribe it to a statistical fluctuation and retain as best-fit fiducial model the one without the Gaussian component. 

\begin{table*}[t]
\begin{center}
\caption{Best-fit parameters derived from geometry-dependent models.}
\label{table:3}
\begin{tabular}{lcccccc} \hline\hline \\
Model & C-stat/d.o.f. & $\rm N_{H}$ & Flux$_{0.5-2}$& Flux$_{2-10}$ & L$_{2-10}$\\ 
&  & {\tiny(10$^{24}$\textit{\rm cm$^{-2}$})} & {\tiny(10$^{-16}$\textit{\rm erg cm$^{-2}$ s$^{-1}$})} & {\tiny(10$^{-15}$\textit{\rm erg cm$^{-2}$ s$^{-1}$})} & {\tiny(10$^{45}$\textit{\rm erg s$^{-1}$})} \\ \hline\\
\textsc{BorSphere$^{(a)}$} & 58.3/48 & ${1.3}_{-0.4}^{+0.4}$  & ${7.1}_{-0.6}^{+5.8}$ & ${5.0}_{-1.4}^{+0.5}$ & ${1.4}_{-0.3}^{+0.4}$ \\
\textsc{Borus$^{(b)}$} & 54.6/48 & ${0.9}_{-0.2}^{+0.2}$ & ${4.4}_{-1.0}^{+0.9}$ & ${6.2}_{-1.0}^{+0.8}$ & ${1.3}_{-0.3}^{+0.4}$ \\ 
\textsc{MYTorus$^{(c)}$} & 48.6/48 & $>$ 6.9 & ${4.3}_{-0.3}^{+2.0}$& ${10.0}_{-1.2}^{+1.2}$$^{(d)}$ & ${>10.0}$$^{(e)}$\\
\\ \hline\hline
\end{tabular}
\end{center}
{\footnotesize $^{(a)}$\textsc{tbabs(BorSphere+ztbabs*cabs*zcutoffpl)}; $^{(b)}$\textsc{tbabs(Borus+ztbabs*cabs*zcutoffpl)}; $^{(c)}$\textsc{tbabs(zpow*MYTZ+MYTS+MYTL)}; \footnotesize $^{(d)}$ The $\rm N_{H}$ is frozen to the best-fit value during the flux calculation; \footnotesize $^{(e)}$ The value reflects the lower limit on $\rm N_{H}$.}
\end{table*}

The derived best-fit column density is $\rm N_{H} \sim 1.4 \times10^{24}\,\rm{cm^{-2}}$. However, it is not strongly constrained and by enlarging the search for a best-fit $\rm N_{H}$ to the entire parameter space including the  $\rm N_{H}\!=\!10^{25} \rm cm^{-2}$ hard limit for the \textsc{MYTorus} model, we obtain a better best-fit (C-stat/d.o.f.= 48.6/48) with $\rm N_{H}$ pegged to $10^{25} \rm cm^{-2}$. In this case, we can only measure a lower limit of $6.9 \times 10^{24} \,\rm cm^{-2}$. Given the very high $\rm N_{H}$, we estimate in 1.2 $\pm$ 0.9 keV the equivalent width relative to the Fe K${\alpha}$ line by substituting the MYTL component with an unresolved Gaussian line at the fixed energy of 6.4 keV and leaving the other \textsc{MYTorus} components at their best-fit values. Table \ref{table:3} reports the derived parameters for this model.

\begin{figure}[h!]
 \centering
{\includegraphics[width=0.49\textwidth] {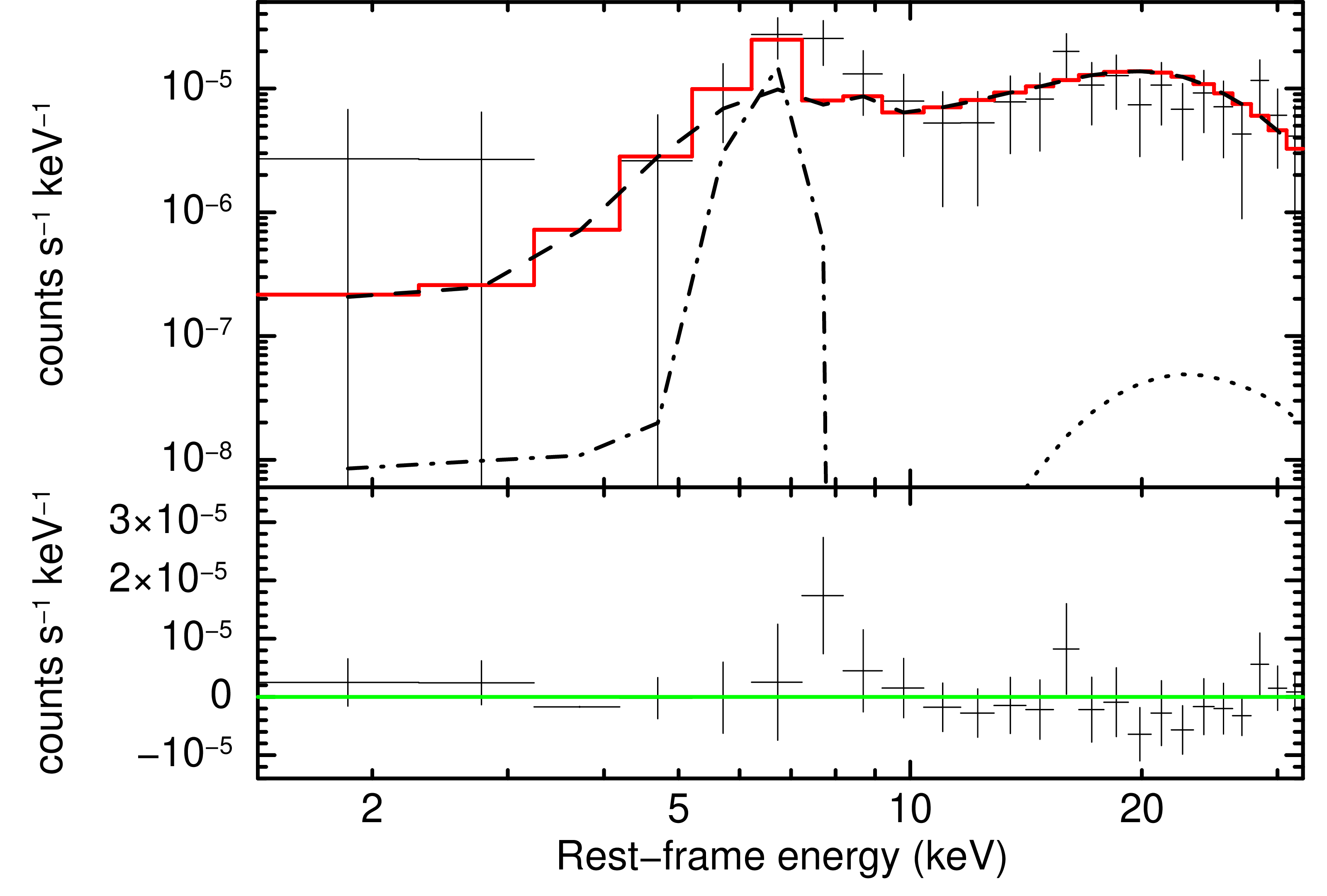}}
\caption{Chandra/ACIS-S spectrum with \textsc{MYTorus} model and residuals. The red solid line represents the best-fit model, the dashed line indicates the absorbed component, the dotted line represents the Compton-scattered component and the dot-dashed line denotes the lines component. The spectrum has been slightly rebinned for better visualization.}
\label{mytorus}
\end{figure}


\section{X-ray emission from the companions overdensity}
\label{X-ray emission companions}
\citet{ginolfi2022} reported the discovery of a significant overdensity of Ly$\alpha$ emitters (LAEs) surrounding the hyper-luminous QSO W0410\,$-$\,09, suggesting the presence of a highly star-forming and dense environment. The LAEs are spectroscopically confirmed through the detection of the Ly$\alpha$ emission line, which provides reliable redshift measurements. The median redshift of the sources is consistent with that of the Hot DOG, suggesting that W0410\,$-$\,09 dominates the environment and lies at the center of the halo’s gravitational potential well. The Ly$\alpha$ luminosities lie in the range $\log\, (\rm L_{{Ly}\alpha}\,[\rm{erg\,s}^{-1}])$ $\approx$ 41.8\,$-$\,42.65 and the star formation rates (SFRs) are $\approx$ 12\,$-$\,100\,$\rm M_{\odot}\, yr^{-1}$.

To search for possible X-ray counterparts of the detected LAEs, we first examined the output of the detection algorithm in the broad, soft and hard bands (see Sect. \ref{reduction}). Excluding the Hot DOG we report the detection of two nearby X-ray sources in the broad and hard bands at a distance from the W0410\,$-$\,09 of $\gtrsim 7~ \rm arcsec$. They are $\gtrsim$ 2.5 arcsec far from their nearest Ly$\alpha$ emitter (Fig. \ref{companions}). Given the Chandra absolute pointing position accuracy and the arcsec-level point spread function, it is highly unlikely that these sources are the high-energy counterparts to the LAEs. The non detection in the X-ray of the LAEs implies that if AGN are present in some of them, they are expected to have X-ray fluxes fainter than the detection limit of our observations. They can either be low luminosity unobscured AGN or higher luminosity obscured AGN. To check for the presence of these undetected AGN, we perform photometry in the broad, soft and hard bands at the position of the LAEs. 
 To this aim we consider only the 19 emitters associated to the Hot DOG with velocities along the line of sight in the range [-2000, +2000] $\rm km\, s^{-1}$ relative to the Hot DOG rest-frame (they are reported in Fig. \ref{companions} as thick crosses). \\
 
 For the photometry, we adopt circular regions of 3 arcsec radius centered on the position of the associated LAEs. We exclude the W0410\,$-$\,09 contribution in the central crowded region by removing the events falling within 3 arcsec radius. We construct a single, combined region that encompasses the entire sky area occupied by all the LAEs, carefully accounting for overlapping areas, which are included only once. We do not detect significant emission in either of the three bands. We further perform photometry in the narrow rest-frame energy band 6\,$-$\,7 keV (observed 1.3\,$-$\,1.5 keV) in order to check for possible Fe K$\alpha$ emission from obscured AGN hosted by the LAEs. A high equivalent width line at 6.4 keV is typically considered the hallmark of heavily obscured sources (see Sect. \ref{empirici}). We find an excess of counts which is $\sim 3 \rm \sigma$ significant over the background emission and corresponding to $9.2_{-3.8}^{+4.5}$ net-counts. For this reason, we run the detection algorithm in the 6\,$-$\,7 keV band and we only find emission from the Hot DOG. This evidence strongly point to undetected obscured AGN emission hosted in several of the LAEs. \\
 
 To estimate a luminosity upper limit for these AGN we adopt the hard band upper limit (the most constraining among the considered bands) which is 24.6 net-counts and corresponds to a count-rate of $8.9\times 10^{-5}\rm \,cts\, s^{-1}$. Assuming a \textsc{Borus} model with $\rm N_{H}=5\times10^{23} \rm \,cm^{-2}$ ($10^{24} \rm \,cm^{-2}$) and normalizing it to this count-rate, we obtain an upper limit for the 2\,$-$\,10 keV unabsorbed luminosity of $ 10^{46} \rm \,erg \,s^{-1}$ ($1.9\times 10^{46} \rm \,erg \,s^{-1}$) which corresponds to a luminosity $ \lesssim 7 \times 10^{44} \rm \, erg \,s^{-1}$ ($10^{45} \rm \, erg \,s^{-1}$) if we assume that all the LAEs host an AGN. \\
 
 We finally estimate the presence of possible thermal diffuse emission permeating the overdense protocluster region. We performed photometry in the three bands from a circular region with radius $\sim 30 \rm \,arcsec$ centered on the Hot DOG. During this procedure we exclude 3 arcsec radius circular regions around the companions and the X-ray detected sources. We do not find significant emission in excess over the background. Given the low Chandra sensitivity at low energy, we further restricted the full and soft bands to energies $> 0.5 \rm \,keV$ (i.e. removing the
0.3\,$-$\,0.5 keV range which can likely contribute with background-only emission). No significant diffuse
emission is found in this case either. Therefore we conclude that there is no detectable diffuse emission in the
overdense region around W0410\,$-$\,09. Either there is no emission, or it is too weak or its
temperature is too low ($\ll$ 1 keV) to be detectable by Chandra.

\begin{figure}[h!]
 \centering
\includegraphics[width=0.48\textwidth] {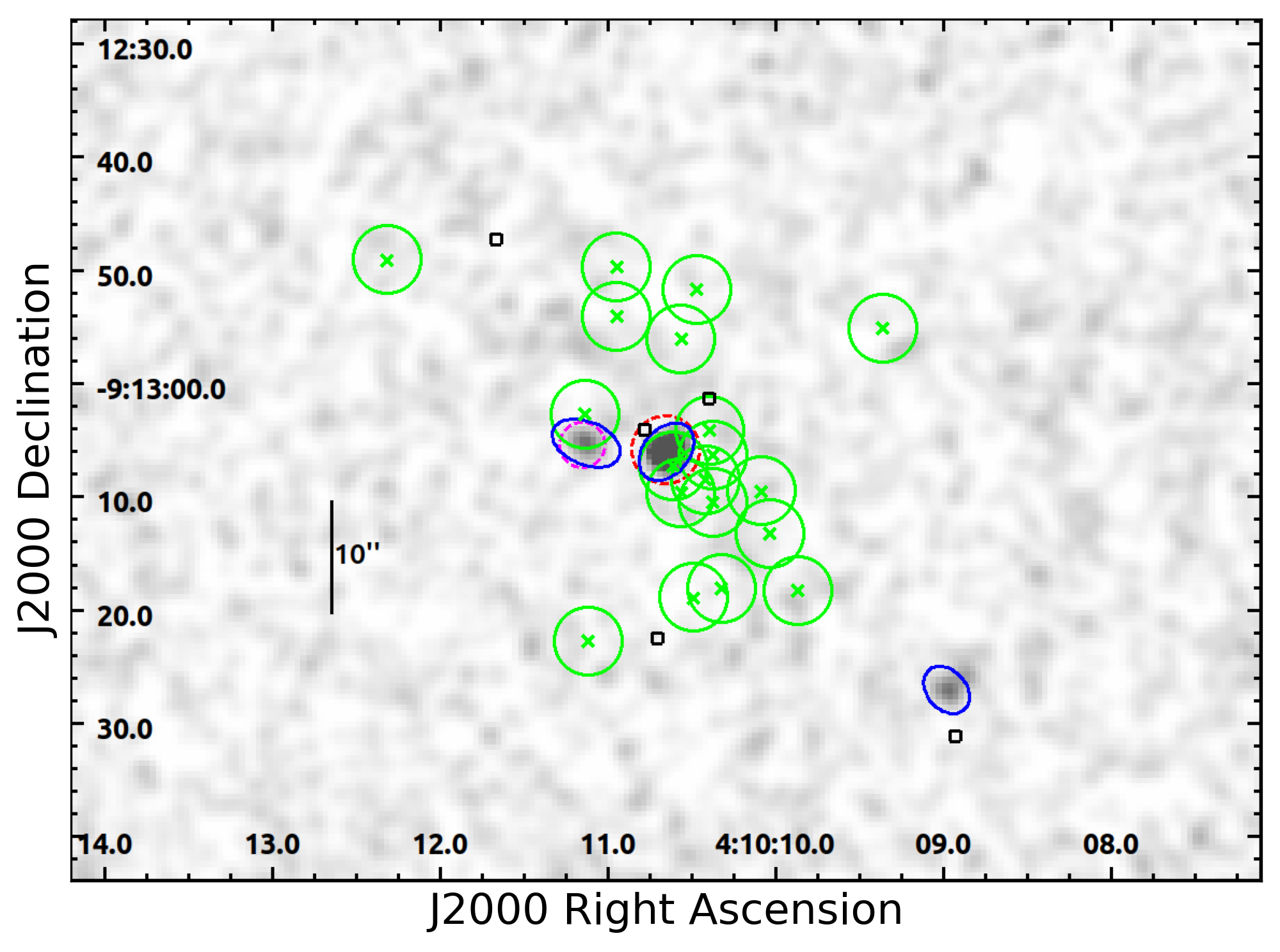}
\caption{ACIS-S image in the 0.3\,$-$\,7 keV energy band showing the 24 LAEs. The 19 companions associated with the Hot DOG are highlighted with green crosses, while those not associated are indicated with black squares. The green circles mark the regions used for the companions' photometry. The X-ray sources detected using the \textit{wavdetect} algorithm are shown in blue. The dashed circles indicate the regions excluded from the companions' photometry, while the Hot DOG is shown in red.}

\label{companions}
\end{figure}


\section{Discussion}\label{discussion}

\subsection{High obscuration and bolometric correction }
Our analysis confirms that W0410\,$-$\,09 is a luminous and heavily obscured QSO shining at  z\,$\sim$\,3.6. An empirical absorbed power-law parametrization accounting for Compton scattering suggests heavy obscuration compatible with CT levels. A similarly good description of its spectrum is a reflection-dominated model further suggesting the source to be obscured at CT levels. Physically motivated models implementing toroidal or spherical geometry for the obscurer, properly accounting for Compton-scattering geometrical effects, still indicate a column density of \(\gtrsim\!10^{24}\, \rm cm^{-2}\). Depending on the models adopted we obtain absorptions from nearly CT ($\sim 10^{24} \rm cm^{-2}$, \textsc{Borus} model) to heavy CT ($\gg 10^{24} \rm cm^{-2}$, \textsc{MYTorus} model) levels. Such a large difference in the derived N$\rm _{H}$ values is likely due to the different geometrical assumptions about the obscuring material in the two models.
Notice that these values are obtained for toroidal obscurers with similar geometry and configuration relative to the observer, i.e. $\rm \theta_{tor}=60 \rm \,deg$ and $\rm \theta_{inc}=80 \rm \,deg$, respectively. The $\textsc{Borus}$ model allows to change $\rm \theta_{tor}$ hence we allow it to vary in order to explore the parameter space. We obtain a best-fit model (C-stat/d.o.f. = 46.6/47) with $\rm \theta_{tor} = 74.3_{-3.4}^{+0.9} \rm \, deg $, $\rm N_{H}\! > \!2.5 \times 10^{24} \rm \, cm^{-2}$ and $\rm L_{x}\!>\! 5.4\times 10^{45} \rm erg \,s^{-1}$. Hence in this case, despite the tight constraints of $\rm \theta_{tor}$, we have a loose constraint on $\rm N_{H}$ still indicating a CT obscurer.

Given its extremely high bolometric luminosity of $6.4\times 10^{47}\, \rm erg \,s^{-1}$ \citep{diaz2021}, W0410\,$-$\,09 is among the most luminous and obscured $\rm z > 3$ AGN ever reported so far as shown by the $\rm N_H \rm \,vs\, \rm L_{bol}$ plot of highly luminous AGN ($\rm L_{bol} > 10^{47} \,\rm erg \,s^{-1}$, see Fig. \ref{nh_lbol}). The bolometric luminosity of the target is conservatively calculated by \citet{tsai2015} by integrating the photometric data with a power-law interpolated between observed flux density measurements.

We estimated a $\rm L_{2-10}$ range from $\sim \!(1.3-1.4)\times10^{45}\rm erg\, s^{-1}$ (\textsc{Borus}) to $>\!10^{46}\,\rm erg \,s^{-1}$ (\textsc{MYTorus}). From these values we calculate $\rm K_{bol,x}=\rm L_{bol}/\rm L_{2-10}$, which is defined as the conversion factor used to estimate the bolometric luminosity from the X-ray luminosity. 
 We obtain the following X-ray bolometric corrections $\rm K_{bol,x}^{Borus}\simeq500$ and $\rm K_{bol,x}^{MYTorus}\lesssim60$.
Fig. \ref{kbol} illustrates the recent $\rm K_{bol}$ vs $\rm L_{bol}$ relationship for Type-II AGN derived by \citet{Duras2020}. The $\rm K_{bol,x}$ of\, W0410\,$-$\,09 for the two toroidal models is reported.
From the relationship calibrated by \citet{Duras2020}, we obtain for W0410\,$-$\,09 a $\rm K_{bol,x}\approx390$, which is consistent with the parameters estimated with the \textsc{Borus} models. The \textsc{MYTorus} modeling would imply an extremely high $\rm L_{2\,-\,10}$ and a very low $\rm K_{bol}$ value (see Fig. \ref{kbol}) which have been rarely reported for hyper-luminous QSOs at Cosmic Noon so far \citep[e.g.,][]{stern2015, martocchia2017, lansbury2020, zappacosta20}. In addition, the $\rm L_{2-10 \,keV}~vs~L_{6 \mu m}$ plot reported in Fig.~\ref{l6lx} shows that the $\rm L_{2-10}$ derived by \textsc{Borus} parametrization agrees with the location of other hyper-luminous Hot DOGs/QSOs and with the best-fit relations reported in the literature. The location for the \textsc{MYTorus} modeling is at least almost more than one order of magnitude in disagreement with both the data and the empirical relations. In order to bring it in agreement we would need to have at least one order of magnitude higher $\rm L_{6\mu m}$.

\begin{figure}[h!]
 \centering
{\includegraphics[width=0.49\textwidth] {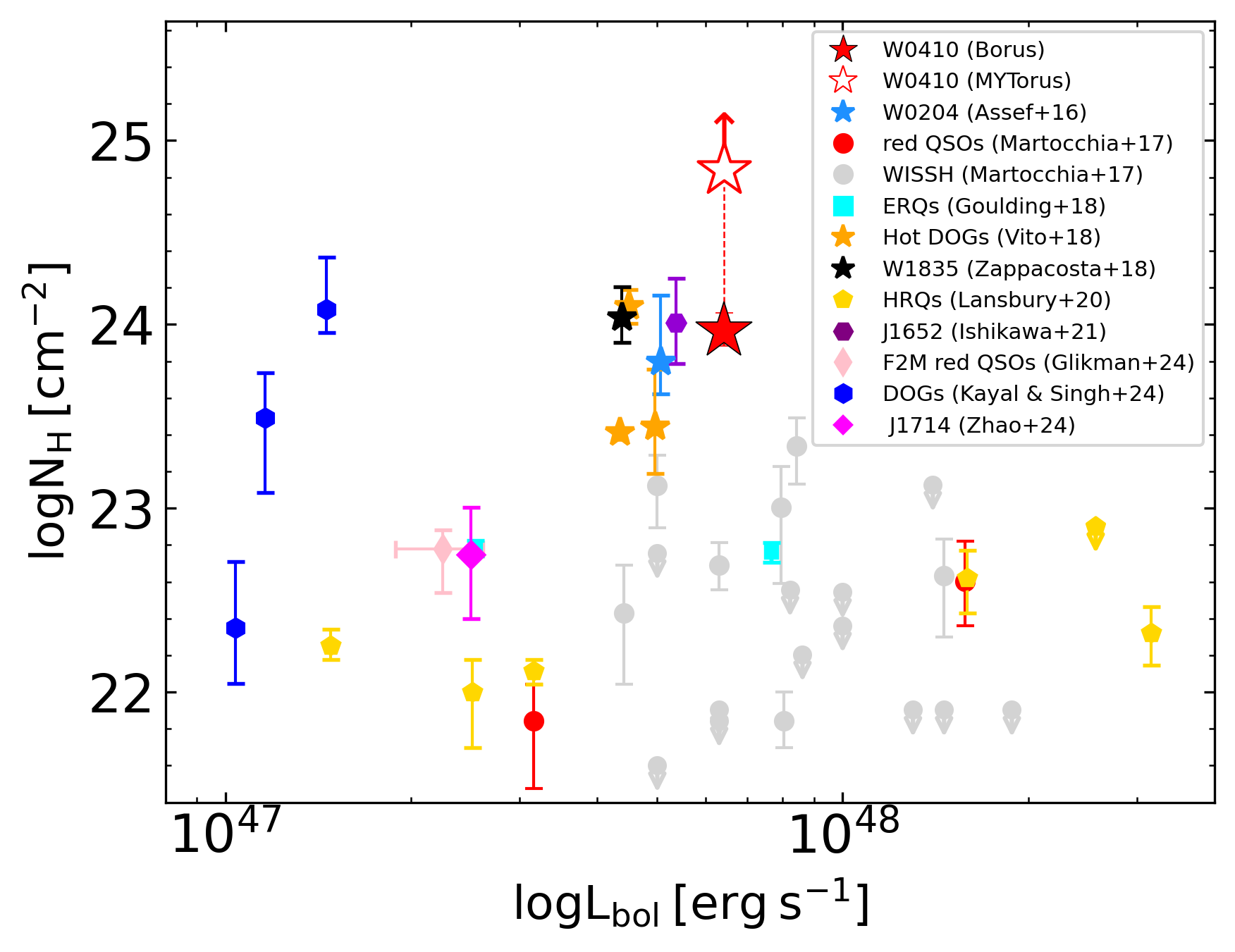}}
\caption{Spectroscopically derived $\rm N_H$ vs $\rm L_{bol}$ for a sample of luminous and X-ray obscured QSOs with $\rm L_{bol}\!>\!10^{47} erg \, s^{-1}$. Filled and empty red stars indicate the \textsc{Borus} and \textsc{MYTorus} parametrizations, respectively. The Hot DOGs are represented with stars of different colors: blue for \citet{assef2016}, orange for \citet{vito2018} and black for \citet{zappacosta2018}.  } 
\label{nh_lbol}
\end{figure}

\begin{figure}[h!]
 \centering
{\includegraphics[width=0.49\textwidth] {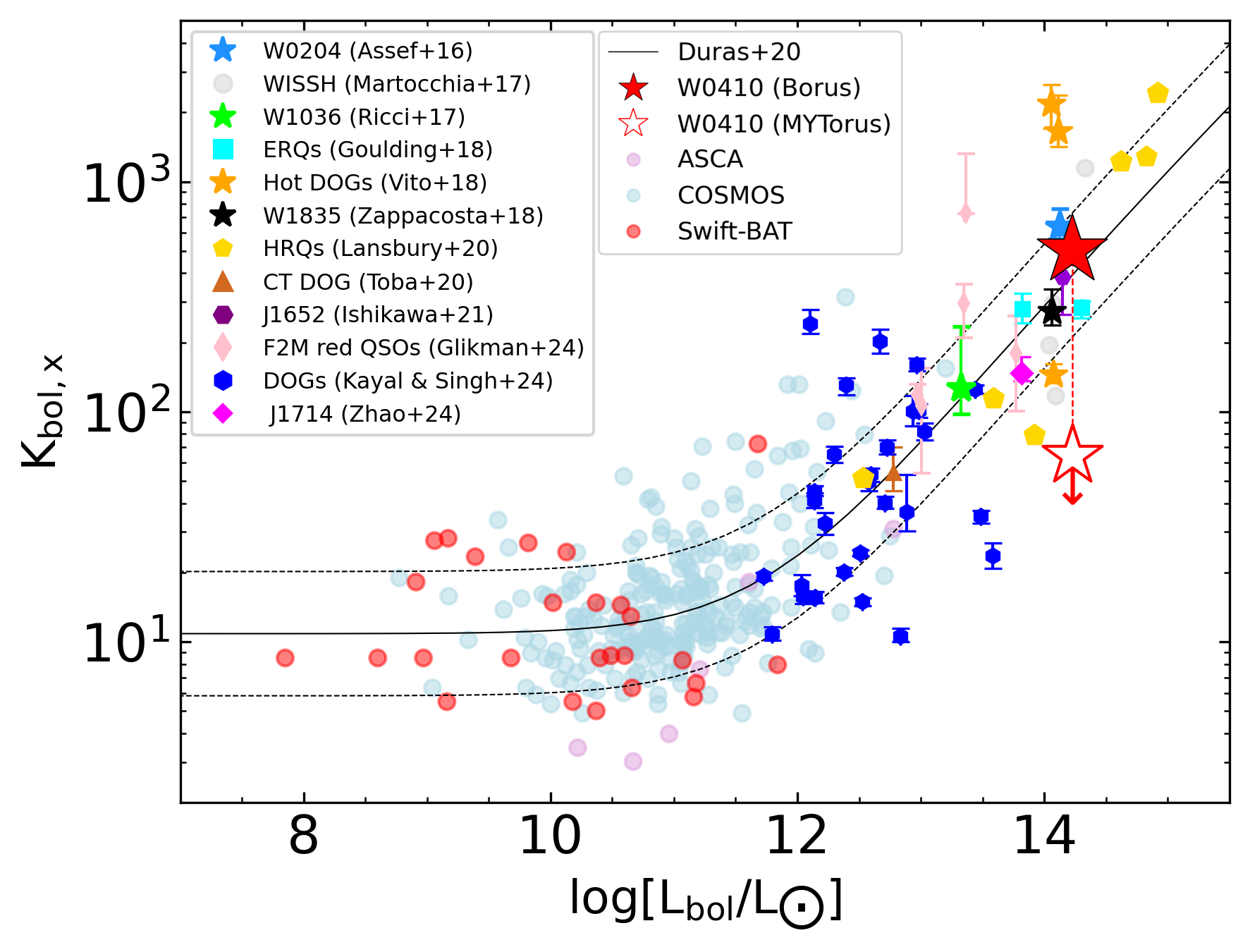}}
\caption{Hard X-ray bolometric correction band as a function of the $\rm L_{bol}$ for Type-II AGN. Filled and empty red stars indicate the \textsc{Borus} and \textsc{MYTorus} parametrizations, respectively, compared to other samples. In particular, we include only the WISSH QSOs with $\rm N_H \!>\! 10^{23} \rm \,cm^{-2}$ \citep[grey circles;][]{martocchia2017}. The light black lines show the best-fit (continuous line) and dispersion (dashed lines) of the $\rm K_{bol,x}-\rm L_{bol}$ relation for Type-II QSOs from \citet{Duras2020}. We also plot the ASCA, COSMOS, and Swift-BAT samples, which include exclusively Type-II AGN.}
\label{kbol}
\end{figure}

 \begin{figure}[h!]
 \centering
{\includegraphics[width=0.49\textwidth] {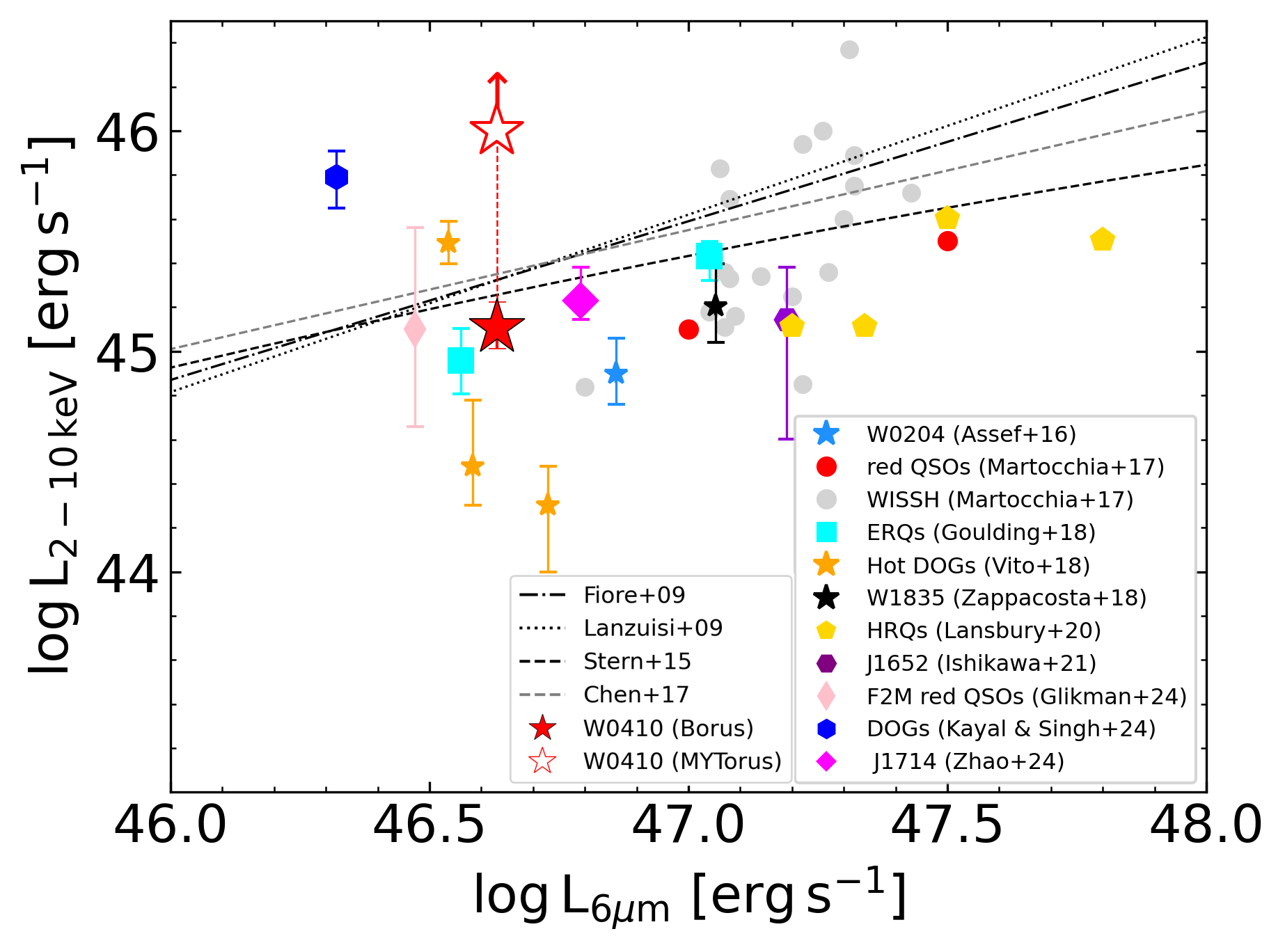}}
\caption{$\rm L_{2-10 \,\rm keV}$ vs $\rm L_{6 \mu m}$  relation for QSOs with $\rm L_{bol}\!>\!10^{47} \rm erg\,s^{-1}$. Filled and empty red stars represent W0410\,$-$\,09
for the \textsc{Borus} and \textsc{MYTorus} models, respectively. The Hot DOGs are represented with stars of different colors as in Fig. \ref{nh_lbol}. Grey points and red dots represent the X-WISSH hyper-luminous Type-I QSO sample \citep{martocchia2017} and two reddened QSOs \citep{martocchia2017}. We also report X-ray-to-MIR relations derived for different optical/MIR/X-ray selected AGN samples \citep{fiore2009, Lanzuisi2009, stern2015, chen2017}.}
\label{l6lx}
\end{figure}

\subsection{The blow-out phase and the circum-galactic nebula}
\label{nebula}

Hot DOGs, being Type-II QSOs, are not expected to show broad lines from virialized nuclear regions. Therefore, in principle, SMBH masses and the inferred Eddington ratios (indicative of their accretion rates) could not be estimated with broad emission lines via single epoch virial mass estimators. Despite this, broad lines have been reported in several Hot DOGs. They are often measured to have large blueshifts indicative of
unvirialized motions perhaps due to nuclear outflows \citep[e.g.,][]{finnerty2020, jun2020, ginolfi2022}, likely leading to biased mass estimates on these sources.
Nonetheless masses around $10^{9}-10^{10}\,\rm M_{\odot}$ have been measured via Balmer, Mg II or C IV emission lines \citep {wu2018,tsai2018,li2024}. 
In particular, \citet{li2024} measured the SMBH masses for a peculiar type of Hot DOGs showing blue-excess emission in the UV-optical band, exceeding the starburst component from the host galaxy. This is consistent with the spectral energy distribution of Type-I sources, being possibly originated by AGN emission scattered outside the obscuring nuclear material \citep{assef2016,assef2020}. They obtained masses in the range $10^{8.7}-10^{10}\, \rm M_{\odot}$. These estimates are based on the C IV emission line, which could be affected by non virial outflowing components, and therefore the corresponding SMBH masses should be overestimated \citep{baskin2005,sulentic2007,denney2012, coatman2017, vietri2018}.

The same calculation onto the more traditional Hot DOGs gave estimates on roughly the same mass range, indicating that, if Hot DOGs and their blue-excess variant are the same sources, we can roughly trust the estimated masses on the reported broad lines for standard Hot DOGs. In this case, the reported mass range is consistent to the mass range measured for hyper-luminous Type-I QSOs \citep[e.g.,][]{vietri2018,Trefoloni2023} at Cosmic Noon.
This provides a first order indication that Hot DOGs are accreting close to the Eddington rate, with Eddington ratios of $\rm \lambda_{Edd}\approx0.1-4$ \citep{li2024}.\\
Fig. \ref{lambdaedd} reports the location of the Hot DOGs in the $\rm N_{H}$ vs $\rm \lambda_{Edd}$ plane. We find that Hot DOGs are located in the upper level of the so called blow-out region. A source populating this region is subject to winds originating by strong radiative pressure on the nuclear dusty gas \citep{fabian2006, fabian2008} which will eventually clears out the surrounding region. This is predicted within the framework of the radiation-regulated unification model \citep[e.g.,][]{jun2021, toba2022, ricci2023}. In this model, AGN dynamically evolve through the $\rm N_{\rm H}$ vs $\rm\lambda_{\rm Edd}$ plane during their life cycle, transitioning from obscured to unobscured phases under the influence of winds. The presence of highly blueshifted broad UV
lines, as reported by \citet{ginolfi2022}, further
corroborates the Hot DOG's blow-out phase. We report W0410\,$-$\,09 at $\rm \lambda_{Edd}=1$ and indicate the $\rm \lambda_{Edd}$ range corresponding to the SMBH mass range reported by \citet{li2024}. Similarly to other Hot DOGs, the source is located entirely in the blow-out region. It is important to note that most SMBH mass estimates used to calculate $\rm \lambda_{Edd}$ rely FWHM of the CIV emission line, which could be affected by outflows (i.e. overestimated SMBH masses and therefore lower $\rm\lambda_{\rm Edd}$). As a result, the true $\lambda_{\rm Edd}$ values may be higher, and the sources would shift further to the right in the $\rm N_{\rm H}$ versus $\lambda_{\rm Edd}$ plane.} Along with the Hot DOGs, we report for comparison, also hyper-luminous high redshift reddened QSOs and local low luminosity AGN. The fact that we observe a nebula significantly smaller than the average size of nebulae reported for Type-I QSOs \citep[e.g.,][]{borisova2016,arrigoni2019,cai2019,fossati2021} and that the nuclear emission is obscured by a star formation medium suggests that the ionizing flux (both X-ray and, especially, UV) is likely blocked and unable to produce an extended nebula, as it arrives strongly attenuated or extinguished on CG scales. 
This scenario is supported by both observational and theoretical evidence regarding CGLANs around these systems. Indeed, \citet{gonzalezlobos2023} demonstrates that dustier systems have smaller CGLANs compared to Type-I QSOs. These observational findings align with theoretical predictions by \citet{costa2022}, who showed that nebula emission from obscured sources (in their case edge-on views) appears to be weak and therefore smaller than expected for a face-on source with similar AGN luminosity. So at fixed bolometric luminosity, looking at the system edge-on or face-on gives a faint or bright nebula, respectively.

Consistent with this scenario, W0410\,$-$\,09 exhibits a CGLAN with a spatial extent of approximately 30 kpc \citep{ginolfi2022}, which is about a factor of three smaller than the typical sizes observed around unobscured quasars, generally reaching $\sim$ 100 kpc (\citealp{borisova2016, arrigoni2019, travascio2020} and references therein).
 Ideally, in the case of an obscuration having a geometry close to 4$\pi$, as expected from the chaotic accretion scenario predicted by the merger-driven QSO formation model, the fact that we observe a nebula suggests some leaked UV emission powering the nebula, implying that the covering factor of the obscuring CT medium could be $<$ 1.

\begin{figure}[h!]
 \centering
{\includegraphics[width=0.49\textwidth] {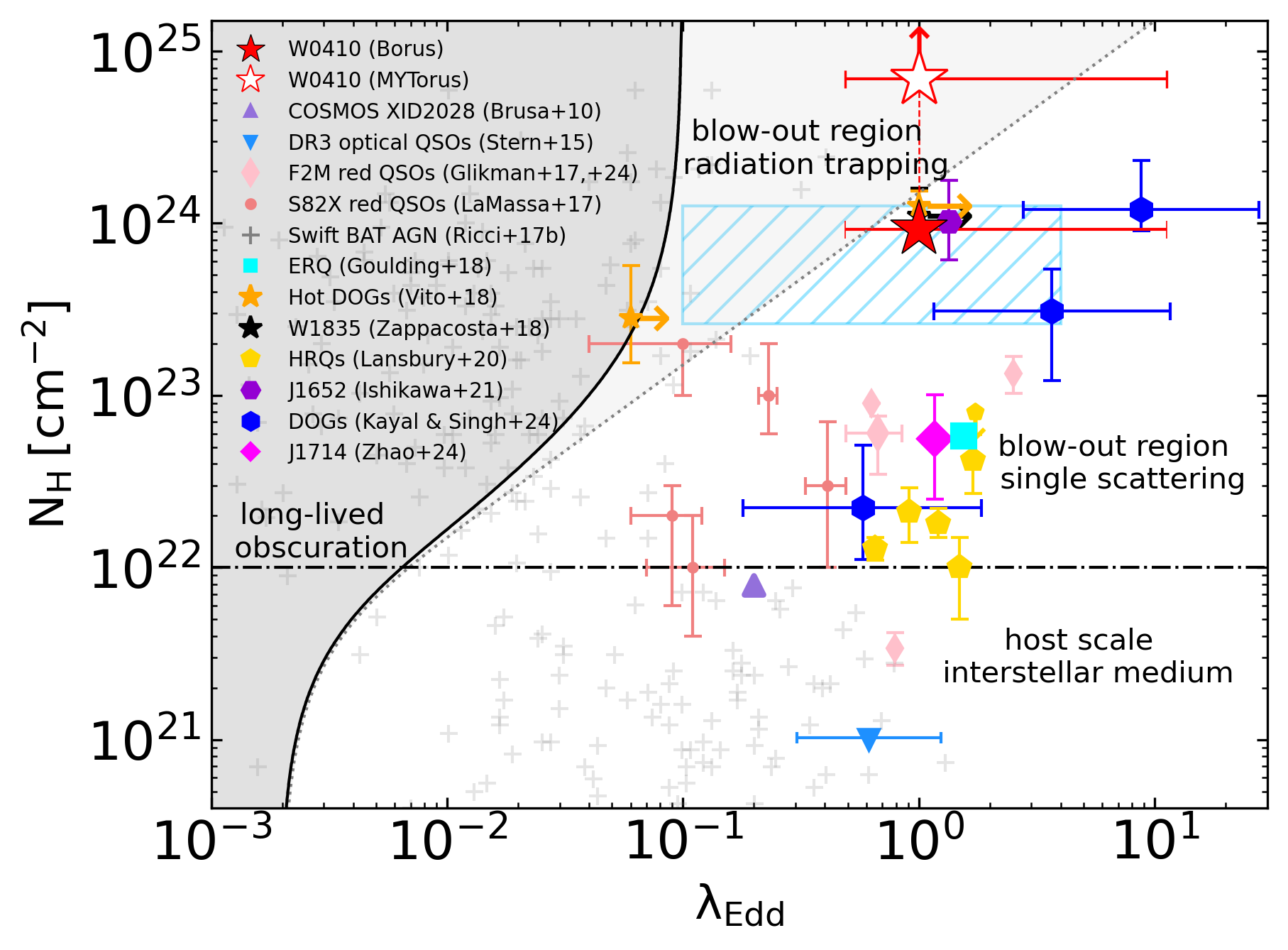}}
\caption{$\rm N_{H}$ vs $\rm \lambda_{Edd}$. We compare only sources with $\rm N_H$ derived from spectral analysis. We report as colored points the luminous QSOs and in grey Swift/BAT low-luminosity AGN \citep{ricciswift}. Filled and empty red stars represent W0410\,$-$\,09 for the \textsc{Borus} and \textsc{MYTorus} models, respectively. The hatched light blue region represents the range of $\rm N_{H}$ values derived from previous Hot DOG X-ray analyses \citep{assef2016, vito2018, zappacosta2018} while the $\rm \lambda_{Edd}$ range was determined using the BH masses estimated by \citet{li2024} (see Sect. \ref{nebula} for details). For hyper-luminous optical QSOs, we show $\pm 1 \rm \sigma$ range of $\rm \lambda_{Edd}$ values for the most luminous QSOs at $\rm z\gtrsim2$ in SDSS DR3 (downward-pointing light blue triangle marker, plotted at $\rm N_{H} = 10^{21} \rm cm^{-2}$). The cross symbols show local ($\rm z\! \sim\! 0.037$) Swift-BAT AGN \citep{ricciswift}. The light gray zone represents the blow-out region for radiation trapping, where the radiation emitted by the black hole is trapped in the surrounding gas through a process of repeated absorption and re-emission, causing continuous heating and accelerating the gas outward \citep{ishibashi2018}. The white zone indicates the blow-out region in the single scattering approximation. Below the dot-dashed line separate nuclear from the host-scale absorption.}
\label{lambdaedd}
\end{figure}

\subsection{Environment}
\begin{figure}[h!]
 \centering
{\includegraphics[width=0.49\textwidth] {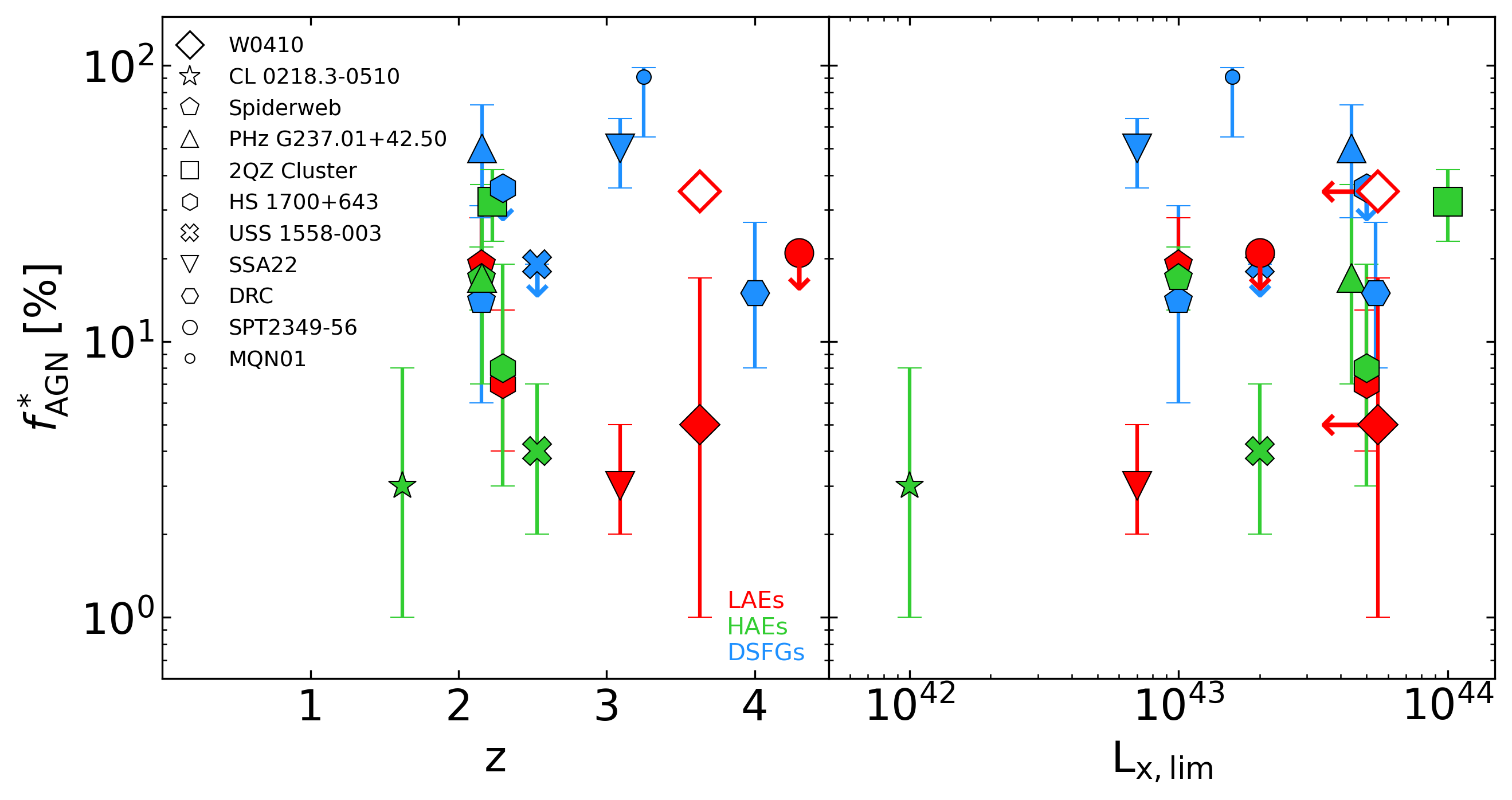}}
\caption{Left panel: Fraction of X-ray-selected AGN among galaxies with different selection criteria in dense environment for high-z overdensities, as reported in the literature \citep{Lehmer2009, digby2010, lehmer2013, krishnan2017, macuga2019, vito2020, polletta2021, tozzi2022, vito2024, travascio2025}. Right panel: X-ray luminosity limit versus AGN fraction in dense environments for high-redshift overdensities. Our measurements are shown as red diamonds: the filled diamond represents the fraction of AGN directly detected among the LAEs, while the empty one refers to the unresolved AGN fraction inferred from stacking analysis at the 90\% confidence level. The colors of the points in the graph indicate the different selection criteria: red for LAEs, green for H$\alpha$-emitting galaxies (HEAs), blue for sub-mm selected dusty star-forming galaxies (DSFGs). Notice that the $f_{\mathrm{AGN}}^*$ of MQN01 is calculated including M$_*\!>\!10^{10.5}$ M$_\odot$ galaxies \citep{travascio2025}.}
\label{fraction}
\end{figure}

The identification of Ly$\alpha$ emitting companions in the UV rest-frame analysis of the VLT/MUSE field of W0410\,$-$\,09 by \citet{ginolfi2022} prompted us to perform a more detailed investigation of their nuclear properties taking advantage of the exquisite Chandra spatial resolution in our deep observation. Within the W0410\,$-$\,09 environment, our analysis does not detect significant X-ray emission coincident with the 19 LAE companions. W0410\,$-$\,09 represents the sole LAE in the field exhibiting X-ray emission. Hence, from the Hot DOG alone we estimated an AGN fraction $f_{\mathrm{AGN}}^{\mathrm{LAE}}=5^{+12}_{-4}\%$ where uncertainties have been calculated assuming low number statistics \citep{gehrels1986}. However, the detection of significant emission from the LAE only in the spectral region corresponding to the Fe K$\alpha$ line and not in the broad energy bands strongly suggests that heavily obscured (CT) AGN are hosted in many of the LAEs. We perform a crude estimation of the possible number of LAEs hosting an obscured AGN by lowering the detection threshold to 90\% level (i.e. selecting LAEs with $\geq2$ X-ray counts in the 6\,$-$\,7 keV rest-frame). Under this assumption we identified 6 sources as “detected”. Adding these unresolved sources to the AGN fraction calculation we find roughly $f_{\mathrm{AGN,u}}^{\mathrm{LAE}}\!\sim\!35 \%$. Fig. \ref{fraction} (left panel) reports the AGN fraction ($f_{\mathrm{AGN}}^*$), measured for distinct galaxy populations
selected adopting different tracers, as a function of redshift for high-z protocluster overdensities. Our $f_{\mathrm{AGN}}^{\mathrm{LAE}}$ estimate is consistent within 
the uncertainties with the $f_{\mathrm{AGN}}^{\mathrm{LAE}}$ values reported in the literature and ranging from 2\% to 19\% \citep{Lehmer2009, digby2010, tozzi2022, vito2024}. In any case, we expect $ f_{\mathrm{AGN}}^*$ to be function of the limiting X-ray flux of the surveyed field. Therefore, we calculate the X-ray luminosity limit ($\rm L_{x,lim}$) reached by our observation of the W0410\,$-$\,09 overdensity. We estimate a 0.5\,$-$\,7 keV count-rate limit of $6.3 \times 10^{-5}$ (90\% confidence level). By assuming a power-law with $\Gamma=1.9$ absorbed by a Galactic $\rm N_{H}=4.03\, \times\, 10^{20} \rm\, cm^{-2}$, we estimate a 0.5\,$-$\,7 keV flux of $8.9 \times 10^{-16} \,\rm erg\, cm^{-2} s^{-1}$, which corresponds at the redshift of the Hot DOG to an unabsorbed (i.e. intrinsic) 2\,$-$\,10 keV luminosity limit of $< 5.5\times 10^{43} \,\rm erg \rm \,s^{-1}$.
Fig. \ref{fraction} (right panel) shows $f_{\mathrm{AGN}}^{\mathrm{LAE}}$ as a function of $\rm L_{x,lim}$ for the protocluster overdense regions targeted in X-rays. We do not find any clear trend between $f_{\mathrm{AGN}}^{\mathrm{LAE}}$ and $\rm L_{x,lim}$ and cannot confirm that AGN activity in LAEs is low in both field and protoclusters as reported in previous works \citep[e.g.,][]{malhotra2003,zheng2016}. Similar fractions and lack of trends are reported also for the $f_{\mathrm{AGN}}^*$ measured in other galaxy populations selected by different tracers (see Fig. \ref{fraction}). Notice that to perform a proper comparison among the AGN fractions the surveyed area volume needs to be taken into account. Furthermore, there are indications that $f_{\mathrm{AGN}}^{\mathrm{LAE}}$ depend on the luminosities of the LAEs \citep{nilsson2011}. In this comparison these corrections have not been applied or tested. Further investigations of AGN fraction in our field with different tracers including also H$\alpha$ emitters,
Lyman break galaxies, sub-mm galaxies,
as reported, e.g., by \citet{vito2024}, are needed to better estimate the AGN fraction in the Hot DOG environment.
This will help in understanding the relevance of different
mechanisms, such as galaxy interactions \citep{ehlert2015}, ram pressure \citep{poggianti2017},
cold gas accretion \citep{gaspari2015}, likely responsible for triggering AGN activity at different scales in these dense environments.

\section{Conclusion}\label{conclusions}
We present the X-ray analysis of $\sim\!280 \rm \,ks$ Chandra observation targeting W0410\,$-$\,09 and its environment. The Hot DOG is significantly detected with $\sim$ 74 net-counts in the full band (0.3\,$-$\,7 keV), making W0410\,$-$\,09 the most significantly detected
Hot DOG in the X-rays. We extract its spectrum and perform
extensive X-ray spectral modeling. The X-ray spectrum is flat and dominated by
a clear emission feature at $\sim$ 7 keV rest-frame, which are signatures of a heavily-
obscured X-ray emission. For this analysis we used both phenomenological and
physically motivated models for the nuclear obscuration. The latter accounts for
the geometry of the obscurer and for an accurate treatment of the X-ray reprocessing
due to Compton reflection. We explored two scenarios: 1) an edge-on
torus, and 2) a sphere isotropically covering the nucleus. We find that:
\begin{itemize}
    \item In all cases, a significant degree of obscuration is measured. The column densities range from nearly CT (\(\log [\rm N_{\rm H}/\rm cm^{-2}] \sim 24\)) to heavy CT (\(\log [\rm N_{\rm H}/\rm cm^{-2}] \gg 24\)) levels. This makes W0410\,$-$\,09 one of the most obscured (if not the most obscured) and luminous \(\rm z > 3.5\) AGN X-ray detected so far;
     \item We measure the intrinsic luminosity in hard band (between 2\,$-$\,10 keV) for W0410\,$-$\,09, finding it to be \(\rm L_{2-10}\!\sim (1.3-1.7) \times 10^{45}\, \rm erg\, s^{-1}\). This agrees with standard relations reported for $\rm K_{bol,x} \rm vs \,\,\rm L_{bol}$ and $ \rm L_{2-10 \,\rm keV}\, vs\, L_{6\mu m}$;
    \item Given the likely high Eddington ratio reported for Hot DOGs by previous works, this level of obscuration suggests that W0410\,$-$\,09 is undergoing a blow-out phase in which the strong pressure of the UV disk emission on the dusty nuclear obscurer is able to overcome the SMBH gravitational pull by blowing-out the circum-nuclear dusty gas as a wind.
   
\end{itemize}
 
 Unlike hyper-luminous unobscured QSOs at Cosmic Noon, which exhibit giant (100 kpc) CGLANs around them, our heavily obscured QSO shows only a small 30 kpc nebula, suggesting that the heavy nuclear obscuration primarily blocks the radiative and mechanical energy transport to CG scales. This extreme obscuration is in agreement with the merger-driven QSO formation scenario, in which our Hot DOG is in the blow-out phase, a crucial stage in the transition from an obscured to an unobscured QSO. This leads to a scenario where 4$\pi$-distributed non homogeneous (i.e. covering factor $<$\,1) heavy absorption of nuclear ionizing radiation limits the powering of the Ly$\alpha$ nebular emission to a few tens of kpc.

Additionally, we perform an analysis of X-ray emission from the surrounding environment. A previous study by \citet{ginolfi2022} revealed that W0410\,$-$\,09 is surrounded by a dense concentration of 19 Ly$\alpha$ emitting companion galaxies located within a CG region of $\sim\!300 $ kpc. This characterizes the Hot DOG neighborhood as one of the densest protocluster environments reported so far.
We do not detect any obvious X-ray emitting counterpart of the 19 LAEs associated to the Hot DOG. Hence, we perform photometry measuring the emission from all the emitters. No emission is detected in the broad,
soft and hard band. However, we report significant (3$\sigma$) emission around 6\,$-$\,7 keV (rest-frame), likely due to a prominent Fe K$\alpha$ line emission, indicating the
presence of heavily obscured AGN hosted by several LAE. We estimate in $5^{+12}_{-4} \%$ the AGN fraction from all the LAEs in the overdensity. This value can reach $\sim35\%$ once accounted for the presence of unresolved obscured AGN hosted by several LAEs. This is in agreement with previous estimates for LAEs in other protoclusters at similar redshifts.

Our results highlight the importance of deep X-ray observations of distant, heavily obscured QSOs for improving our comprehension of the complex phenomenon
of the assembly and evolution of the most massive galaxies. This study can be
considered a pathfinder for future investigations of the nuclear properties of high-z dust-enshrouded QSOs, which constitute one of the main objectives of the new X-ray telescopes
currently in the design phase.

\begin{acknowledgements}
The authors acknowledge financial support from the Bando Ricerca Fondamentale INAF 2022 Large Grant “Toward an holistic view of the Titans: multi-band observations of z > 6 QSOs powered by greedy supermassive black holes”.
LZ acknowledge support from the European Union – Next Generation EU, PRIN/MUR 2022 2022TKPB2P – BIG-z. FR acknowledges financial support from the Italian Ministry for University and Research, through the grant PNRR-M4C2-I1.1-PRIN 2022-PE9-SEAWIND: Super-Eddington Accretion: Wind, INflow and Disk-F53D23001250006-NextGenerationEU. SC gratefully acknowledges support from the European Research Council (ERC) under the European Union’s Horizon 2020 Research and Innovation programme grant agreement No 864361. FV acknowledges support from the “INAF Ricerca Fondamentale 2023 – Large GO” grant. We thank Andrea Travascio for providing the MQN01 values included in Fig. \ref{fraction}.
\end{acknowledgements}

\bibliographystyle{aa}
\bibliography{bibliografia}

\end{document}